\documentclass[12pt,preprint]{aastex}

\def\arcsec{{\prime\prime}}  

\shorttitle{The Role of Magnetic Topology}
\shortauthors{Lee et al.}

\begin{document}

\title{The Role of Magnetic Topology in the Heating of Active Region Coronal Loops}


\author{J.-Y. Lee\altaffilmark{1,2}, Graham Barnes\altaffilmark{2},
K.D. Leka\altaffilmark{2}, Katharine K. Reeves\altaffilmark{1},
K. E. Korreck\altaffilmark{1}, 
L. Golub\altaffilmark{1}, and E. E. DeLuca\altaffilmark{1} } 
\altaffiltext{1}{Harvard-Smithsonian Center for Astrophysics, 
Cambridge, MA 02138} 
\altaffiltext{2}{NorthWest Research
Associates, CoRA Division, Boulder, CO 80301}

\begin{abstract}

We investigate the evolution of coronal loop emission in the context
of the coronal magnetic field topology. New modeling techniques allow
us to investigate the magnetic field structure and energy release in
active regions. Using these models and high resolution
multi-wavelength coronal observations from the {\it Transition Region
and Coronal Explorer} ({\it TRACE}) and the X-ray Telescope (XRT) on
{\it Hinode}, we are able to establish a relationship between the
light curves of coronal loops and their associated magnetic topologies
for NOAA Active Region 10963.  We examine loops that show both
transient and steady emission, and we find that loops that show many
transient brightenings are located in domains associated with a
high number of separators.  This topology provides an environment for
continual impulsive heating events through magnetic reconnection at
the separators.  A loop with relatively constant X-ray and EUV
emission, on the other hand, is located in domains that are not
associated with separators.  This result implies that larger-scale
magnetic field reconnections are not involved in heating plasma in
these regions, and the heating in these loops must come from another
mechanism, such as small-scale reconnections (i.e., nanoflares) or wave
heating.  Additionally, we find that loops that undergo repeated
transient brightenings are associated with separators that have enhanced
free energy. In contrast, we find one case of an isolated transient
brightening that seems to be associated with separators with a smaller
free energy.

\end{abstract}
 

\section{Introduction}

Observational studies of the character and evolution of coronal
structures are becoming more refined as observational capabilities
improve and accompanying models by which to interpret the results are
developed. In studies of the evolution of coronal emission,
observations at coronal-emission wavelengths show a variety of
structures that evolve in distinct ways.  It is useful to examine
these observations in the context of the magnetic field.
Coronal loops observed in soft X-ray and EUV have been studied since
the Skylab observations in the 1970s \citep[e.g.,][]{vaiana73,
yoshi95, berger99}. In previous studies, the spatial and temporal
correlations between soft X-ray observations and EUV observations of
coronal loops have been investigated in order to study coronal loop
evolution \citep{schmi04, nitta00, nagata03}.  \citet{wineb05} found
that structures observed in the X-rays and then subsequently in the
EUV are due to emission from heated coronal loops cooling through the
passband of each instrument.

Recent studies using the {\it Hinode}/EUV imaging spectrometer (EIS)
have presented observations that show a brightening in EUV following a
brightening in X-rays \citep{warren07, urra09, warren09a,
warren09b}. A multi-thread hydrodynamic simulation has reproduced
several coronal loop characteristics, including high electron density
and a long observed life time, for the coronal loops near 1\,MK, but
it has difficulty reproducing the high temperature loop emission
observed by XRT \citep{warren09a}. It has also been shown that simple
impulsive and quasi-steady heating does not reproduce the observed
loop characteristics \citep{ugarte06}.

Several studies have shown evidence for magnetic field changes during
the coronal loop evolution \citep{wineb05, warren07}. Total magnetic
flux in active regions has been used to investigate if the heating
rate is proportional to a power of the total magnetic flux
\citep{warren06, lund08, fludra08}. Bulk active-region properties can
provide statistical guidance \citep{fisher98}, but as \citet{urra09}
show, different coronal emission lifetimes can be related to the field
strength and its variation at loop boundaries.  One recent study
suggests that the changes in the topology of the magnetic field may be
related with the evolution of the high temperature coronal loop in
X-ray emission \citep{warren09a}.

Starting from magnetogram observations, Magnetic Charge Topology (MCT)
models \citep[see][for an overview]{longcope05} provide a framework to
analyze the coronal topology, including the magnetic field
connectivity and separator field lines. In this class of models, flux
concentrations on the boundary are represented by point charges or
sources.  The magnetic field lines established by the MCT model form
domains, defined as continuous volumes containing field lines that
connect the same pairs of sources \citep{longcopek02}. Thus, the MCT
model can determine which domains are along a given line of sight in a
potential (current-free) extrapolation of the coronal magnetic field.

Domains are bounded by separatrix surfaces, whose intersections form
separator field lines, and thus are the location of reconnection in
MCT models. \citet{priest05} have proposed that slow photospheric
motions can generate electric current near separators and separatrices
which can be released as a heating energy. Recently, \citet{noglik09}
showed that a potential magnetic field extrapolation represents the
structure of the hotter X-ray loops and the larger cool loops seen in
171\,\AA\ images appear to follow the separatrix surfaces determined
by the MCT analysis.  In addition, \citet{plowman09} show that
separators explain the observed uniform width of coronal loops better
than randomly distributed potential field lines.  In this paper we use the MCT
model to determine the location of flux domains and separators. By
construction, this model disallows reconnection, but can identify the location
of the separators where reconnection is most likely to occur.  The details of
how reconnection across the separators heats coronal loops in a flux system
requires a more detailed time dependent MHD simulation that is beyond the scope
of this paper.

The Minimum Current Corona (MCC) model \citep{longcope96} has been
developed in order to determine the quasi-static evolution of magnetic
fields starting with the potential fields of the MCT model and
introducing currents along separators in response to footpoint
motions. The MCC model represents the corona with the minimum
permissible current after the evolution using a series of quasi-static
steps and subject to the constraint that there is no reconnection
\citep{longcope96}. The resulting increase in energy above the
potential field energy (free energy) can be released in the form of
heating \citep{longcope96, priest00, priest05}.  This model has been
applied to several flare events to investigate the energy source of
the flares \citep{longcope07, jardins09a}.

In this paper, we use the MCC model to calculate the energy build-up
from the photospheric magnetic field evolution, presenting this output
in the context of multi-wavelength observations of the temporal
evolution of the coronal emission.  The MCT model is used to establish
potential magnetic field connectivity, determine separators and
domains along the line of sight at various coronal heights. The MCC
model is used to determine coronal currents and quantify the energy
buildup. The results of the MCC analysis are compared to the evolution
of the coronal loop brightness, comparing selected areas of both steady
emission and transient emission.

In \S2, we describe the observational data employed in this
analysis. In \S3, we explain briefly the MCT and MCC models. We also
describe the magnetic topology of AR~10963 as characterized by the MCT
model and the free energy calculated by the MCC model in \S3. In \S4,
we discuss the observed coronal loop brightness evolution as compared
to the results of the MCT and MCC models. In \S5, we present our
conclusions.

\section{Observational Data}
\label{sec:obs}

The target of this study is NOAA Active Region 10963 (AR\,10963),
which transited the visible solar disk from 2007 July 8--23. This
region is a very small, exceptionally quiet numbered active region,
producing a few B- and small C-class flares (peak flux in the {\it
Geostationary Operational Environmental Satellite (GOES)} 1--8\,\AA\
soft X-ray band of $\simeq 10^{-7}, 10^{-6}{\rm\,W\,m}^2$,
respectively). The region was event-quiet above the B-class threshold
from July 11 00:05\,UT until July 18 13:50\,UT.  The day of interest for
this study was during the quiet time, focusing on 2007 July 14.

The X-ray telescope \citep[XRT,][]{golub07} on board the joint
NASA/JAXA mission {\it Hinode} \citep{kosugi07} observes the solar
corona at high ($\sim$\,1$''$ pixels) spatial resolution in multiple
bandpasses.  For this study, we use a sequence of images from the
Al/Poly filter, which samples plasma at $>$ 2\,MK, with a 20\,s
cadence from 18:27\,UT on July 14 to 01:47\,UT on July 15. The XRT
observed AR\,10963 with deep exposures in the C/poly ($>$ 3\,MK)
filter to study faint structures until 18:15\,UT on July 14. These earlier
observations do not allow for study of the changes in coronal
brightness since coronal loop images were saturated.

The {\it Transition Region and Coronal Explorer} \citep[{\it
TRACE},][]{handy99} observations are at a 1 minute cadence in the
171\,\AA\ band, which samples the coronal plasma at $\sim$\,1\,MK with
$\sim$\,0.5$''$ pixels.

The line-of-sight component of the pixel-averaged photospheric
magnetic field was obtained from the Michaelson Doppler Interferometer
(MDI) aboard the {\it Solar and Heliospheric Observatory} \citep[{\it
SoHO},][]{scherrer95}.  For this study, a roughly 24\,hour time
sequence of the 96-minute cadence, $\sim\,2^\arcsec$ data was used,
from 2007 July 14 01:39\,UT to 2007 July 15 00:00\,UT.

\subsection{Co-alignment among XRT, {\it TRACE}, and MDI}

Co-alignment between the datasets arising from three different
instruments on three different spacecraft was crucial for this study,
and was performed in the following manner.

We used the MDI level 1.8 data and full-sun, level 1 data from the
Extreme-ultraviolet Imaging Telescope (EIT) on {\it SoHO} (calibrated
by \textit{eit\_prep.pro} in the SolarSoft suite of analysis
tools). We first supposed that an internal co-alignment between MDI
and EIT is correct, and the EIT images then provided the common
coalignment platform for both the XRT and {\it TRACE} data.  Active
region 10963 was observed by XRT and {\it TRACE} for $\sim$\,6 hours
with Al/Poly (1.0286$^\arcsec$ pixels) and 171\,\AA\ (0.5$^\arcsec$
pixels), respectively. The EIT (2.63$^\arcsec$ pixels) observed the
full sun every 6 hours at 284\,\AA\ and 171\,\AA.

The XRT observed the full solar disk with Ti/poly every 20 minutes as
a context image. X-ray bright points on the full Sun at 19:07:15\,UT on
July 14 were aligned with EIT 284\,\AA\ at 19:06:07\,UT on July 14. Due
to the similar response to high temperature plasma, the EIT 284\,\AA\
and the XRT Ti/poly observations were used to co-align the
satellites. The spacecraft jitter on {\it Hinode} \citep{shimizu07}
was corrected with the reference image at 19:07:15\,UT.

The {\it TRACE} 171\,\AA\ observation at July 14 18:59:26\,UT was
aligned with the EIT 171\,\AA\ observation at July 14 19:00:13\,UT. The
aligned image was used as a reference to align for later time
observations for $\sim$\,6 hours (using
\textit{trace\_cube\_pointing.pro} in SolarSoft) to account for jumps
in the {\it TRACE} pointing as the active region was tracked across
the disk.

\subsection{Area selection for light curves}

\begin{figure} 
\plotone{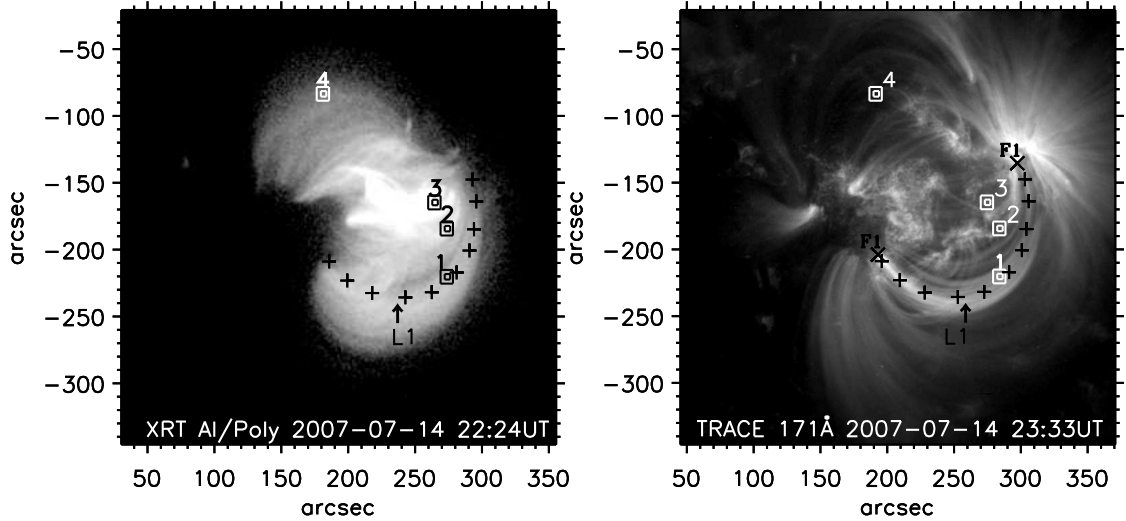}
\caption{The locations of four boxes for light curves
(Figure~\ref{fig:lightcurves}) on XRT (left) and {\it TRACE} (right)
images. Smaller and larger boxes represent the areas of $\sim 5''
\times 5''$ and $\sim 10'' \times 10''$, respectively. A Loop (L1)
that produces the enhancement in light curve Box~1 is represented by
$''+''$ symbols along the loop. The ends of this loop are represented
by F1. This figure is also available as a mpg animation. The movie
shows the XRT and {\it TRACE} observations for $\sim$ 6 hours
corresponding to the light curves in Figure~\ref{fig:lightcurves}.}
\label{fig:xrttrace}
\end{figure}

We select four areas in the XRT and {\it TRACE} data with which to
investigate coronal brightness changes. These areas are presented as
four boxes in Figure~\ref{fig:xrttrace}. Inner and outer boxes
correspond to approximately $5^\arcsec \times 5^\arcsec$ and $10
^\arcsec \times 10^\arcsec$ areas, respectively. The Boxes~1, 2, and 4
were selected based on the coronal loops in the XRT observation, and
Box~3 was selected based on a small loop visible in the {\it TRACE}
data at 22:24\,UT on July 14. 

Light curves for each box are presented in
Figure~\ref{fig:lightcurves}. Box 1 shows a brightening in the {\it
TRACE} observation about an hour and a half after a decrease in X-ray
brightness. Box~2 shows transient brightenings in XRT, however the
{\it TRACE} light curve shows only very small changes in
brightness. Box~3 shows transient brightenings in both XRT and {\it
TRACE}.  As seen in the animation version of
Figure~\ref{fig:xrttrace}, the light curves of Boxes~2 and 3 include
the brightness changes for several loops. Box~4 shows steady emission
in both XRT and {\it TRACE}, at a lower level in XRT than any of the
other boxes.  The light curves in Figure~\ref{fig:lightcurves} show
similar behaviors for the two box sizes, which verifies that the light
curves do not depend on a small change in the location of the box.

Figure~\ref{fig:xrttrace} also shows a loop (L1) that contributes to the light
curves in Box~1 on the XRT and the TRACE observations at 22:24\,UT and
23:48\,UT, respectively, when the brightenings are seen in each
instrument. The ends of the loops seen by the XRT observation are
represented with F1. These locations are used to
estimate the height of the loop in \S4.1 (see also
Table~\ref{tb:loops} in Appendix). The location of the loop for
the Box~1 is selected based on the loop observed by the {\it TRACE}
because the loop is better resolved in the EUV observation, and then
the same location is represented on the XRT observation.

\section{Magnetic Topology and Energy Buildup}
\label{sec:models}

We utilize a Magnetic Charge Topology model, and the Minimum Current
Corona model to understand the magnetic topology and the
magnetic energy changes in the active region that we observe with XRT
and {\it TRACE}.  We use the MCT model to define the magnetic topology and locate
possible sites for magnetic energy reconnection. With the MCC model, we quantify
the magnetic energy changes. The details of these models and our
implementation of them for the analysis of AR\,10963 are described in
the next subsections.

\subsection{Magnetic Charge Topology model}

The MCT model is used to establish the magnetic connectivity by
potential magnetic field extrapolation into the corona from a
distribution of point sources representing the photospheric magnetic
field distribution \citep{bau80,gor88,pri89,lau93,dem94,par94}.  As a
result of the extrapolation, the magnetic field lines form a domain,
defined as a volume containing field lines that have the same sources
at their ends (unless the field lines tend to infinity). In the
following paragraphs, we explain several topological terms derived
from the inferred magnetic connectivity that we use in this paper.

Locations where the magnetic field vanishes are referred to as
magnetic null points. In three dimensional geometry, the local
magnetic structure of a null can be described in terms of a fan
surface and two spine field lines \citep{parnell96}. The fan surface
is a set of field lines which radiate out from a null point (positive,
B-type null) or radiate into a null point (negative, A-type null). The
spine field lines are those directed toward the positive null point or
directed away from the negative null point.

The fan forms a separatrix surface that divides the corona into
volumes of different connectivity called domains.  Separators are
field lines which begin at a positive null and end at a negative null,
and they are located at the intersection of separatrix surfaces.
Magnetic reconnection and energy release in a coronal field stressed by
photospheric motions are likely to occur near separators, although it
should be noted that the MCT model does not allow for reconnection.

The flux in a connection ($\psi_{ij}$) is determined from the number
of magnetic field lines initiated from the source ({\it i}) in random
directions and followed to their termination at a source of the
opposite polarity ({\it j}) \citep{mct}. For each pair of sources, it
is possible to have connections through multiple domains
\citep{longcopek02}. In this analysis, each different pair of sources
defines a single domain. Thus the connectivity flux and the domain
flux are the same for each pair of sources in this work.

\subsection{Partitioning the magnetogram time-series}

To extrapolate the magnetic field lines into the corona, 
the MCT model requires a set of distinct point sources derived from an
observed magnetic field. Therefore, the observed magnetic fields must be
partitioned in a way that represents the surface magnetic flux
concentrations. To determine the flux concentrations from the
photospheric magnetic field observations, the observed magnetic
features are partitioned to preserve important magnetic morphology,
and the location of each source is determined by the flux weighted
center of the partition \citep{mct}.

We use a time series of 96-minute-cadence MDI magnetograms for the
photospheric magnetic field observation. First, we select the region
of AR\,10963 from the full-disk magnetograms for $\sim$\,24 hours,
taking into account solar rotation. The selected area retains the
balance of positive and negative sources within a 4$\%$ ratio of the
total signed to total unsigned magnetic line of sight signal.  The
selected regions are transformed from the image plane (or ``plane of
the sky'') to a plane that is tangent to the Sun's surface at the
flux weighted center of the selected active region.  The ``$\mu$-correction'' is
applied, which assumes that all field is radial, and divides the
line-of-sight magnetogram by the cosine of the observing angle, to
approximate the radial field. This process transforms the observed
line-of-sight pixel-averaged magnetic field component to a tangent
plane approximation of the magnetic flux, as required by the MCT
model.

\begin{figure} \epsscale{0.7}\plotone{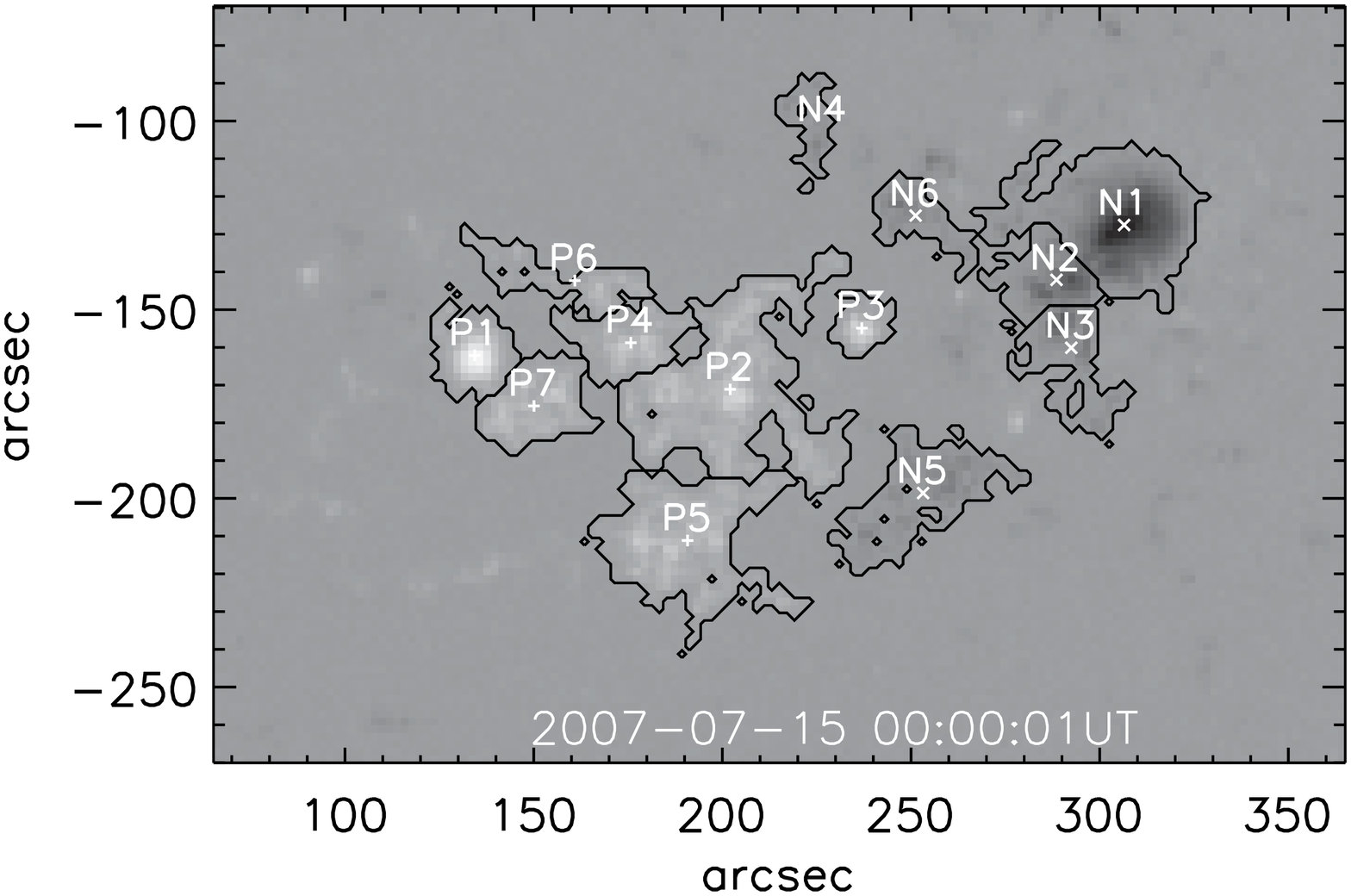}\plotone{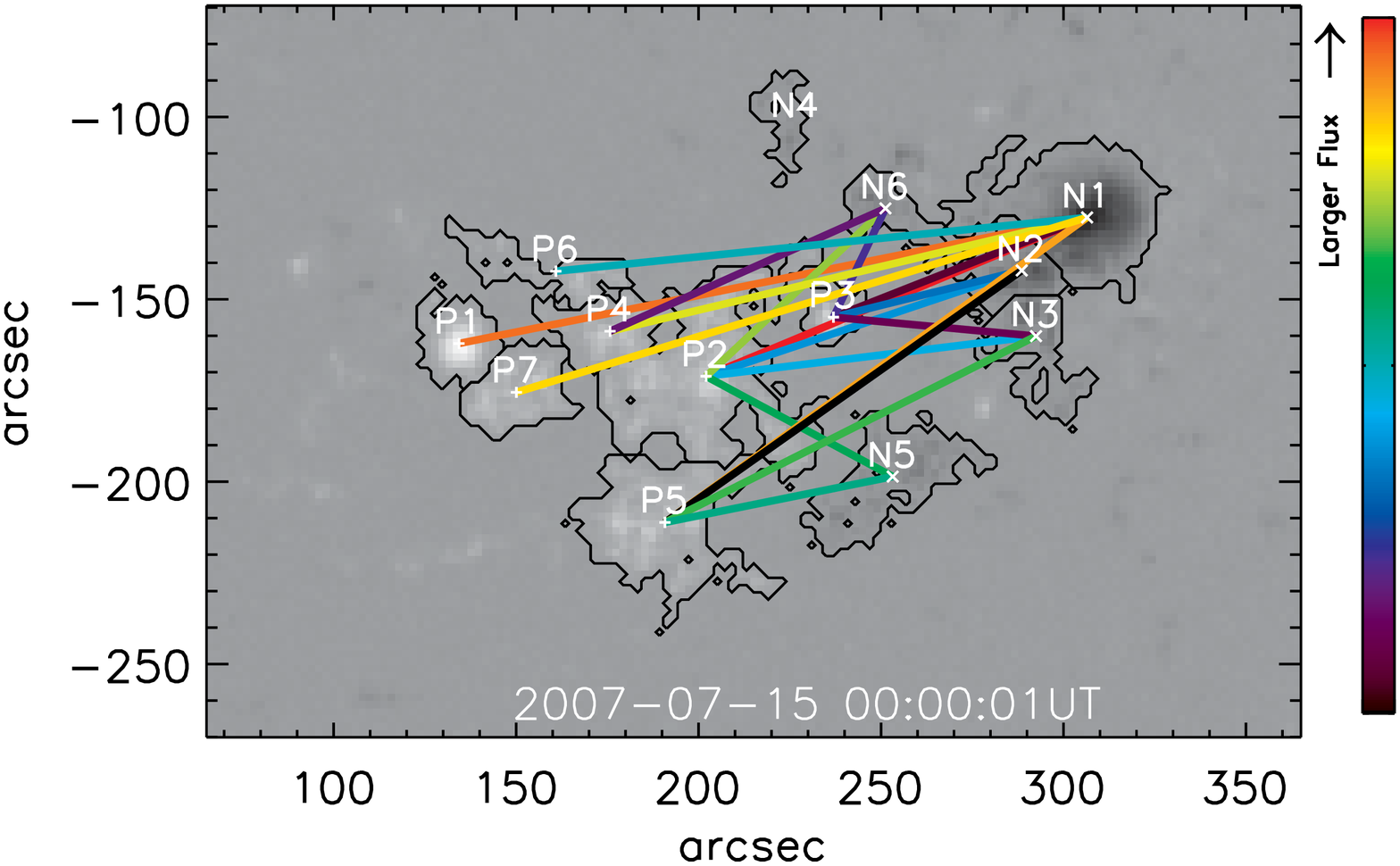}
\caption{Top:~Partitioning at 00:00\,UT on July 15. Pole locations of
positive (P: $+$) and negative (N: $\times$) sources are
represented. Contour represents the partitioning. The background MDI
images show only the smoothed field above the threshold used in the
partitioning algorithm.  Bottom:~Colored lines represent domain fluxes
among all sources.  The contours surrounding individual pixels which
are included or excluded from a partition are rendered as small diamonds.}
\label{fig:ar}
\end{figure}

The top panel of Figure~\ref{fig:ar} shows the boundaries which result
from partitioning the positive and negative sources.  We initially
determine the flux concentration at the last time observed, July 15
00:00\,UT, since an earlier study has found that performing the
partitioning in reverse chronological order provides the most
consistent partitioning in light of noise and solar evolution
\citep{longcope07}.  The partitioning parameters used are similar to
those in Barnes et al. 2005, though they have been adjusted to match
this specific active region\footnote{Specifically, for this dataset we
employ a field threshold of 3 times the detection threshold, a
smoothing depth of 0.5\,Mm, a saddle point merging level of 200\,G,
and a minimum of source flux of $10^4\,{\rm G\,Mm}^2$ \citep[see][for
a description of these parameters]{mct}.}.  Recently,
\citet{longcopebb09} found that there are only slight differences in
connectivity that result from using different partitioning parameters, 
so our results are unlikely to be sensitive to the
choice of partitioning parameters. The bottom panel of
Figure~\ref{fig:ar} shows domain fluxes among all sources at the same
time as the top panel.

MDI 96-minute data consist of images with integration times of 30 and
300 seconds during the 24 hour period of interest, and different integration
times could conceivably change the inferred flux. To mitigate this effect, the
detection level was assigned to be the 3\,$\sigma$ level of a Gaussian fit
for the histogram of each MDI observation. This procedure sets an appropriate 
background level for each integration time. No significant
systematic difference was detected in the resulting fluxes above the
3\,$\sigma$ threshold.

Beginning from the end of the time series, the observed magnetic
elements in the magnetograms are advected to the earlier MDI
observations using Fourier Local Correlation Tracking
\citep[FLCT,][]{flct}, in reverse chronological order for the same
reasons as mentioned above. The FLCT finds the velocities of features
in two successive images by computing the cross-correlation function
using standard Fast Fourier Transform techniques
\citep{Welsch04}\footnote{For the implementation of the FLCT here, we
employ time differences between two images of 1, unit of length of a
single pixel of 1, and sub-images weighted by Gaussian width of 45,
see \cite{flct}.}.  These velocities are horizontal velocities for
each pixel in the MDI observations. The computed velocities are then
used to track elements and determine their partitioning assignments in
subsequent observations. Thus, this procedure preserves the
partitioning information as consistently as possible through the time
series.

\begin{figure} \epsscale{0.7}\plotone{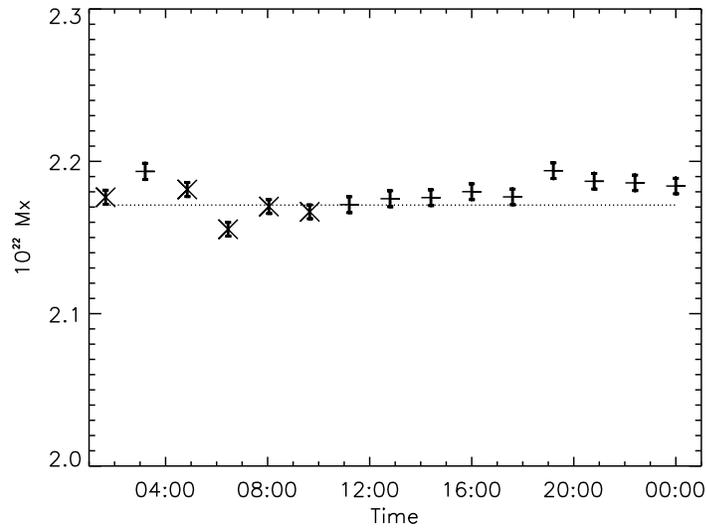}
\caption{Total unsigned source flux for all sources obtained by the
time-series partitioning. The integration times of the MDI
observations are represented with uncertainties in the total unsigned
flux: 300 sec ($\times$) and 30 sec ($+$).  Dotted line represents the
total averaged flux used in this analysis excluding the source flux of
N4 ($\Phi_{N4}=0$, see details in text). Averaged uncertainty for 15
times in the total unsigned flux is $5.0\times10^{19}$\,Mx.}
\label{fig:arflux}
\end{figure}

In Figure~\ref{fig:arflux}, we show a plot of the total unsigned flux for
all sources obtained by the time-series partitioning, and the
time-averaged total flux.  It is clear that there is little variation
during the $\sim$~24 hours, implying that there is no significant
emerging or submerging flux during this time period. Therefore, we
impose that the flux {\it for each source} is also constant in time,
since the flux changes that do occur are too small to contribute
significantly to the active region field. We use the time-averaged
flux for each source in the MCT model. This approach prevents the transfer
of flux between neighboring sources in the partitioning, and precludes
changes in connectivity due to changes in source flux resulting from
emergence or submergence, thus simplifying the usage of the MCC model
\citep{longcope07}, as described in \S3.4.

The source N4 (near [220$^\arcsec$,~-100$^\arcsec$] in
Figure~\ref{fig:ar}) is submerged at 17:36\,UT on July 14, and only
appears in 4 time steps. The average flux of N4 for the four
sequential MDI observations when it is above threshold is $2.6 \times
10^{20}$\,Mx. However the average fluxes of other sources are in the
range of $(4.9-67) \times 10^{20}$\,Mx.  Therefore, we do not consider
the source N4 for this analysis on account of its relatively small
flux and short lifetime. This choice helps the analysis stay
consistent with the assumption that there is no emerging/submerging
flux.

\begin{figure} \epsscale{0.7}\plotone{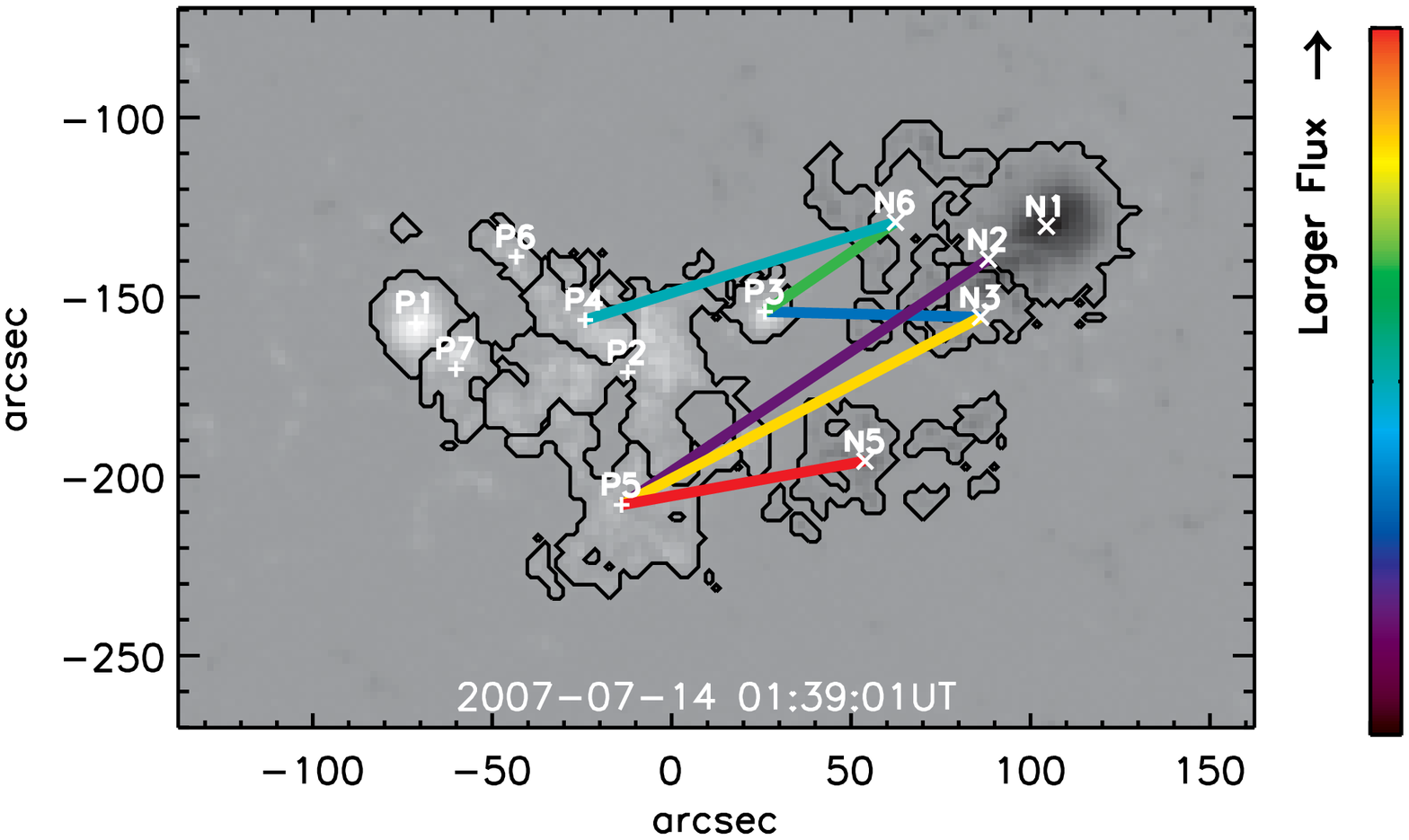}
\epsscale{0.7}\plotone{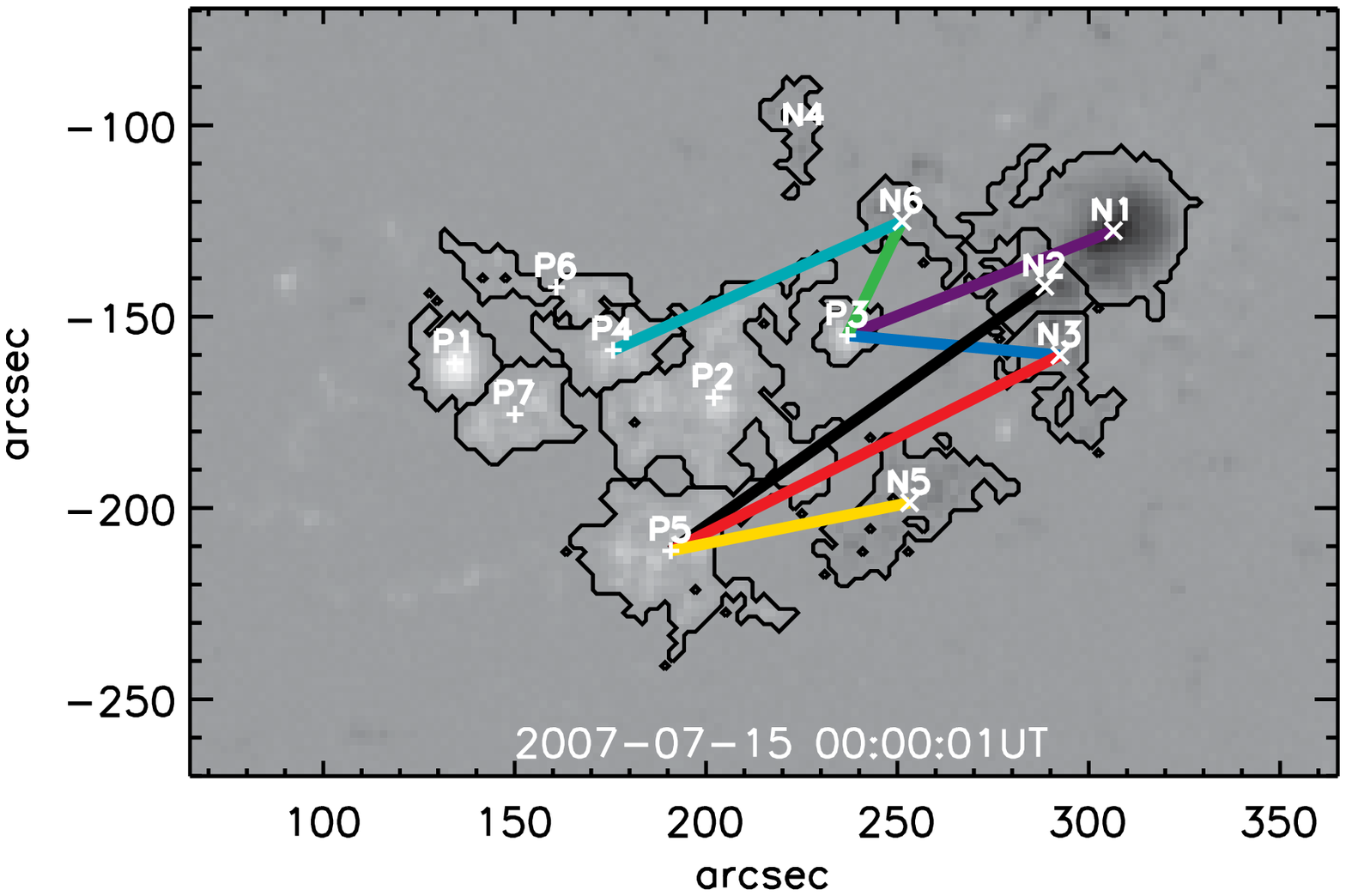}
\caption{Seven domains related to the separators
(Table~\ref{tb:separators}) are shown on the top and the bottom panels
at 01:39\,UT on July 14 and 00:00\,UT on July 15, respectively. Colored
lines from each positive ($+$) and negative ($\times$) source
represent the flux in each connection. The scale of
color is the same in both top and bottom panels, but this scale is
different from the right panel of Figure~\ref{fig:ar}. Footpoint
motions and the changes of domain fluxes over the 24 hour period can be seen 
by comparing the top and the bottom panels.
}
\label{fig:connect}
\end{figure}

In Figure~\ref{fig:connect} we demonstrate the evolution of
partitioning, source locations, and selected connection fluxes
($\psi_{ij}$) calculated by the MCT model as described in \S3.1, for
the first and last time.  Only connections related to the seven
separators discussed at length below are shown.  In this figure, the
change in the source locations between the first time and the last
time can be compared. For instance, it can be seen that P3 and N6 are
approaching with a rotation during $\sim$ 24 hours indicating that
this region is generating significant currents (see details in \S3.4).

\subsection{Magnetic topology of AR\,10963}

The MCT model was used to establish the magnetic topology of AR\,10963 at
00:00\,UT on July 15.  The flux concentrations obtained from the MDI
observations by the partitioning explained in the previous section were used as
a set of distinct sources. Figure~\ref{fig:top} shows  seven of the separators
for this active region, specifically those which are topologically related to
the areas in the light curves shown in Figure~\ref{fig:lightcurves}. 

In order to track the build up of currents along the separators, it is useful
to represent the domain fluxes ($\psi_{ij}$) in terms of the source fluxes
($\Phi_k$) and the separator fluxes ($\Psi_\sigma$), where the indices $i$, $j$
and $k$ refer to sources while the index $\sigma$ refers to separators.
Because all the sources are in the plane of the photosphere, there is a
``mirror corona'' below the photosphere, which is a reflection of the corona.
Thus for each separator, there is a mirror separator, and the two form a closed
curve.  To express the domain fluxes in terms of the source and separator
fluxes, it is convenient to first determine which domains pass through the
closed curves formed by each separator and its mirror.  

To illustrate how to determine which domains pass through such a closed curve,
consider separator 5 (hereafter $\Psi_5$) which starts and ends at the nulls
B11 and A3, respectively, and consists of the intersection of the separatrix
surfaces from these two nulls.  The separatrix surface associated with null A3
intersects the photosphere along the fan trace from A3 to source P4, then along
the spines of null B8 to source P2, then along the spines of null B11 to source
P3, and returns to null A3 along the fan trace from P3. This surface encloses
only source N6, thus all field lines within this separatrix surface must
terminate on N6.  Similarly, the separatrix surface associated with null B11
only encloses source P3, thus all field lines within this separatrix surface
must start on source P3. Since $\Psi_5$ is the intersection of these separatrix
surfaces, the domain P3--N6 passes through the closed curve consisting of
$\Psi_5$ and its mirror image below the photosphere.

Alternatively, one can consider the photospheric footprint of domain P3--N6.
Since all the sources lie in the plane of the photosphere, any field line
originated in this plane must, by symmetry, remain in this plane. One can see
in Figure~\ref{fig:top} that any field line initiated at source P3 in the
photosphere must pass under $\Psi_5$ in order to terminate on source N6, and
hence the domain P3--N6 passes through the closed curve of $\Psi_5$ and its
mirror. 

As a somewhat more complicated example, consider $\Psi_3$, whose endpoints are
on nulls B6 and A2.  The separatrix surface associated with B6 encloses only
source P5, but the separatrix surface associated with A2 encloses both source
N3 and source N5 (Note that source N5 is also enclosed by the separatrix
surface associated with null A1, and this separatrix surface is, in turn,
enclosed by the separatrix surface from null A2, but the important point is
that source N5 is enclosed by the separatrix surface from A2).  Thus, domains
P5--N5 and P5--N3 both pass through the closed curve consisting of $\Psi_3$ and
its mirror.  Once again, one can instead consider the photospheric footprint of
these domains.  Any field lines initiated in the photosphere from P5 must pass
under $\Psi_3$ to reach source N5, and similarly for field lines which
terminate on source N3.  The domains enclosed by the separators and the nulls
where the separators start and end are shown in Table~\ref{tb:separators}. 

In addition to the domains associated with each separator, each null whose fan
is unbroken (that is, each null for which all field lines in the fan surface
end on the same source and thus is not the start/end of a separator; see
\cite{longcopek02}) implies the existence of a domain with flux equal to the
flux of the spine source enclosed by this domain. For example, null B10 has an
unbroken fan, and so the flux in domain P6--N1 equals the flux of source P6.
This constrains those domain fluxes not associated with any separator. 

Knowledge of the domains passing through the closed curves of the separators
and their mirrors, combined with knowledge of the nulls with unbroken fans is
sufficient to express the domain fluxes in terms of the separator and source
fluxes.  The flux in each domain not enclosed by any separator is determined
from conservation of the source fluxes, $\Phi_i = \displaystyle\sum_{j=1}^n
\psi_{ij}$, where $\Phi_i$ is the source flux determined from the observed
magnetic field obtained by the partitioning process, $\psi_{ij}$ is the domain
flux between the $i^{th}$ and $j^{th}$ sources, and $n$ is the total number of
sources of the opposite polarity to source $i$.  The result, shown in
Table~\ref{tb:matrix}, provides information about which domains can be affected
by each separator. This information on the relationship between domain fluxes
and separator fluxes is used to interpret our results in \S4.

\subsection{Minimum Current Corona model}

In the previous section, we established the magnetic topology of
AR\,10963 using the MCT model which assumes that the coronal magnetic
field is a potential (current-free) field. \citet{longcope96} 
developed the MCC model to incorporate source evolution, 
current generation along separators, and energy build-up in the context
of the MCT model. We utilize the MCC model to quantify the magnetic
free energy of AR\,10963 in \S3.5. In this section, we describe briefly the
MCC model \citep{longcope96,longcope01}.

As a constraint in the MCC model, \citet{longcope01} has proposed a
flux-constrained equilibrium (FCE). This equilibrium is found by minimizing the
magnetic energy subject to the constraint that, in the absence of emergence and 
submergence of flux through the boundary, the domain matrix does not change.
This constraint is equivalent to prohibiting reconnection in the corona.  The
minimum energy magnetic field occurs when current flows only along separator
field lines, such that each domain is current-free (potential) but is encircled
by current ribbons that cancel the non-potential flux introduced by footpoint
motions.

In a three dimensional magnetic field, the currents occur at points
topologically equivalent to separators in the potential field
\citep{longcopec96, longcopek02}. Therefore, the current in the
minimum energy state can be found using the flux through a
separator. In the minimum energy state, the flux ($\Psi_\sigma$)
through a separator ($\sigma$) can be expressed as \citep{longcope96} 

\begin{equation} \Psi_\sigma = \Psi_\sigma^{(v)} +
\Psi_\sigma^{(I_\sigma)} + \sum_{\sigma' \ne \sigma}
\Psi_\sigma^{(I_{\sigma'})},
\label{eq:sep}
\end{equation}
where $I_\sigma$ is the current flowing along a separator $\sigma$. 
The vacuum term ($\Psi_\sigma^{(v)}$) is the sum of the domain
fluxes enclosed by the separator in the potential field (see
\S3.3). The second term on the right hand side, ($\Psi_\sigma^{(I_\sigma)}$) is a self-flux
from the current along the separator. The third term is the flux from
other separator currents.

At the first time ($t_0$), we assume that there is no current. Then, from
Equation~\ref{eq:sep} we get
\begin{equation} 
\Psi_{\sigma} (t_0) = \Psi_\sigma^{(v)} (t_0).
\label{eq:t0}
\end{equation}
Using the constraint on the domain flux that there is no
connectivity change, Equation~\ref{eq:sep} becomes, at $t_0$ and later time
$t_1$,
\begin{equation} \Psi_\sigma(t_0) = \Psi_\sigma^{(v)}(t_0) =
\Psi_\sigma(t_1) = \Psi_\sigma^{(v)}(t_1) + \Psi_\sigma^{(I_\sigma)}(t_1)
+ \sum_{\sigma' \ne \sigma} \Psi_\sigma^{(I_{\sigma'})}(t_1).
\label{eq:t1}
\end{equation}
Therefore, the flux difference in the potential field
($\Delta \Psi_{\sigma}^{(v)}$ = $\Psi_{\sigma}^{(v)} (t_1)$ -
$\Psi_{\sigma}^{(v)}(t_0)$) should be canceled by the sum of the
second and the third terms in the right side of Equation~\ref{eq:t1} because
the $\Delta \Psi_{\sigma}$ is zero with no connectivity changes.

In order to determine the current along the separator field line, the
MCC model requires a relationship between the self-flux and
current.  The self-flux produced by current along the separator 
\citep{longcope96, longcopem04} is given by,
\begin{equation} \Psi_{\sigma}^{(I_{\sigma})} = \frac{I_\sigma
L_\sigma}{c} \ln \left( \frac{eI_\sigma^*}{|I_\sigma|} \right)
\label{eq:self}
\end{equation}
where $e$ is the base of the natural logarithm, $I_\sigma$
is the current along separator, $L_\sigma$ is the length of the
separator in potential field, and $I_\sigma^*$ is a characteristic
current proportional to the average perpendicular magnetic shear along
separator (see Longcope \& Magara 2004 for details).

The third term on the right hand side in Equation~\ref{eq:sep}, the flux
from other separator currents, is expressed with a mutual inductance
matrix \citep[$M_{\sigma \sigma'}$,][]{longcope96, longcopeb07},
\begin{equation} \sum_{\sigma' \ne \sigma}\Psi_\sigma^{(I_{\sigma'})}
= \sum_{\sigma' \ne \sigma} M_{\sigma \sigma'} \frac{I_{\sigma'}}{c},
\label{eq:mutual}
\end{equation}
assuming that two separators are far enough apart, so the change in 
the flux enclosed by separator $\sigma'$ will not be greatly affected by the internal
distribution of the current on $\sigma$.
Therefore, the current along the separator field line is
calculated by
\begin{equation} \Delta \Psi_\sigma ^{(v)} = \frac{I_\sigma
L_\sigma}{c} \ln \left( \frac{eI_\sigma^*}{|I_\sigma|} \right) +
\sum_{\sigma' \ne \sigma} M_{\sigma \sigma'} \frac{I_{\sigma'}}{c}.
\label{eq:current}
\end{equation}
The above equation is used to find the current $I_\sigma$, and
thus the energy change for each separator, $\Delta W_{MCC}$, can then 
be calculated from the following
\begin{equation} \Delta W_{MCC} = \frac{1}{4\pi}
\int\limits_{\Psi_\sigma^{(v)}}^{\Psi_\sigma} I_\sigma d \Psi_\sigma =
\frac{L_\sigma I_\sigma^2}{2 c^2} \ln\left(
\frac{\sqrt{e}I_\sigma^*}{|I_\sigma|}\right) + \frac{I_\sigma}{2c^2}
\sum_{\sigma' \ne \sigma} M_{\sigma \sigma'}{I_{\sigma'}}.
\label{eq:energy}
\end{equation}

From the above equations, we evaluate the current along each separator
field line which arises due to footpoint motions, and the inferred
excess energy over the potential field, referred to as the free energy.
This free energy is available to be released in the form of localized heating
\citep[e.g.,][]{longcope01, priest00, priest05}.

\subsection{Magnetic energy of AR\,10963}

We utilize the MCC model to investigate the magnetic energy changes of
AR\,10963. The evaluated magnetic energy changes are compared with the
coronal loop brightness evolution observed by XRT and {\it TRACE} in
\S4.

The MCC model infers the current and the free energy by using the
changes of the potential field flux through the separators as
described in the previous section. However, the magnetic topology
evolves with time, so separators that are present at one time are not
necessarily present at the next \citep[e.g.,][]{jardins09a}. Therefore
one cannot follow the same separator for the entire time series, and a
different approach is needed to determine the change in flux enclosed by each
separator.

The fundamental assumption of the MCC model is that there is no
magnetic reconnection in the corona, thus the flux in each domain must
remain constant.  As described in \S3.3 and shown in
Table~\ref{tb:separators}, we have related the flux enclosed by each
separator to the flux of the set of domains enclosed by the separator.
Therefore, the flux through separators used in the MCC model has been
replaced by the appropriate combination of domain fluxes, smoothed
with a boxcar function with a width of 3 time steps at each time to
reduce the noise due to partitioning.

We evaluate the magnetic energy changes assuming that the magnetic
field at the first time, July 14 01:39 UT, is a potential field with
no current. Then, the domain fluxes are used to determine the changes
in separator flux used in Equation~\ref{eq:current} to derive the
currents, and Equation~\ref{eq:energy} is used to determine the free
energy. In the initial stages, the model does not represent the
coronal currents well, since the system is started in a potential
state, but it should improve as it evolves away from the potential
state. Since the energy evolves slowly over several hours in this active
region, the choice of different initial time within a few hours will
not affect the results of our energy calculation.

\begin{figure} \epsscale{1.1} \plottwo{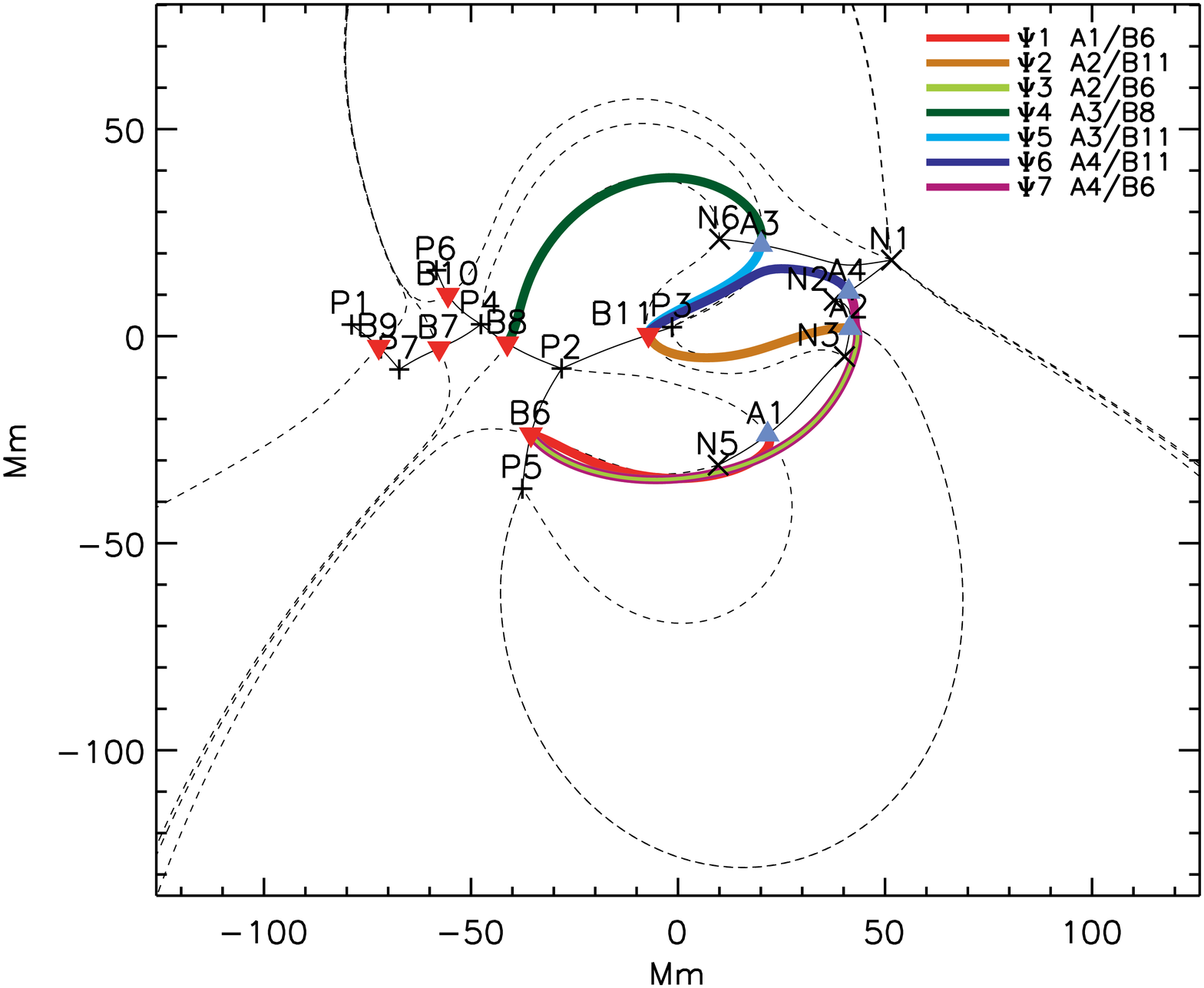}{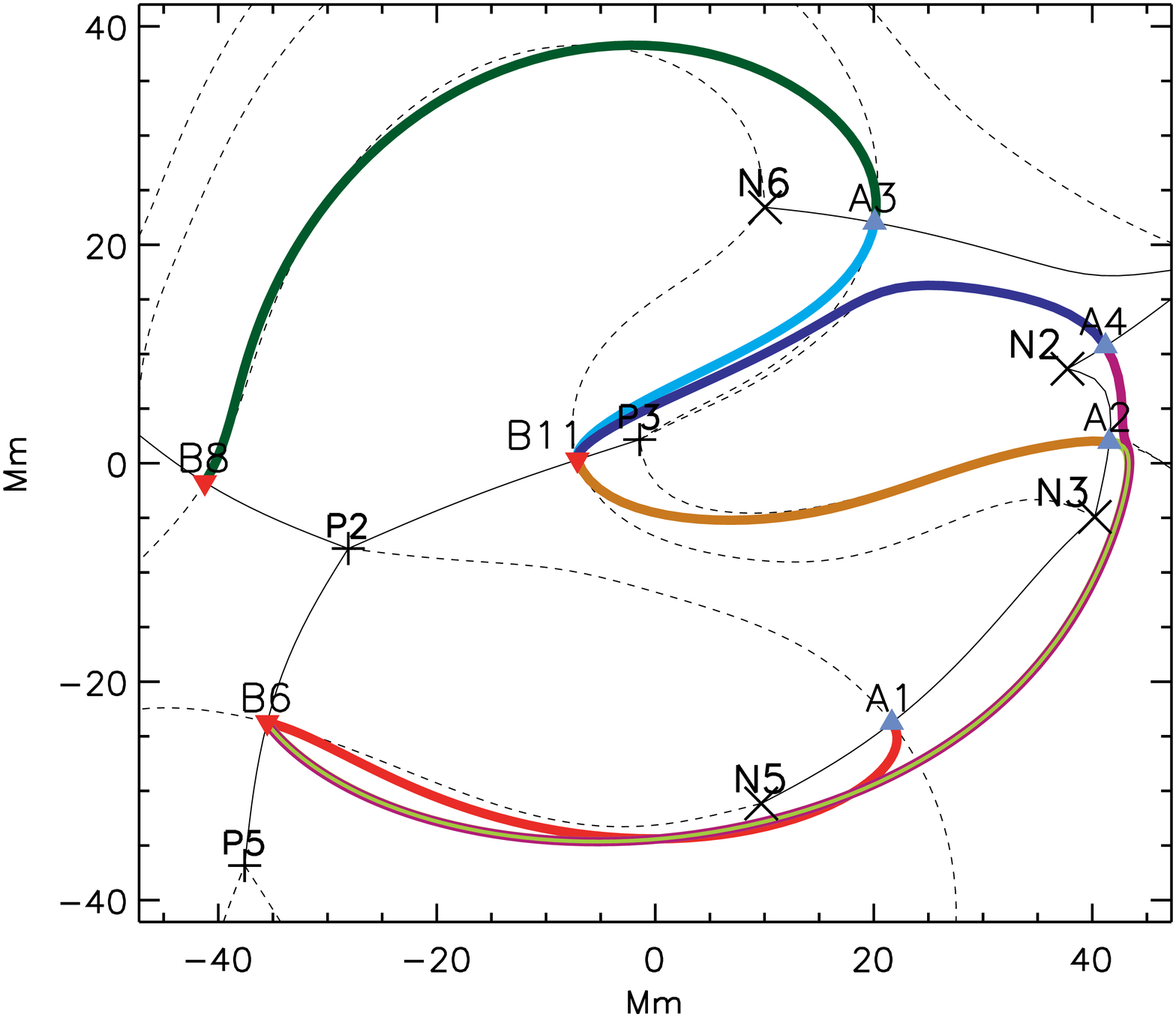}
\caption{Topology of AR\,10963 at 00:00\,UT on July 15. Right panel
shows an enlarged version of the middle of left panel. We show the
topology on the tangent plane to explain the domains enclosed by
separators in \S3.3. Nulls (B-type: red $\bigtriangledown$ and A-type:
blue $\bigtriangleup$) are located between positive and negative
sources, respectively. Black dashed and black thin solid lines are fan
traces and spines that show where separatrix surfaces intersect the
photosphere. Colored lines are the separators: the intersection of
separatrix surfaces. $\Psi_3$ (A2/B6) is represented by a thinner
line than other separators to be visible with $\Psi_7$ (A4/B6) because the
locations of these two separators are close except between nulls
A2 and A4. 
}
\label{fig:top}
\end{figure}

In Figure~\ref{fig:energy} we show the free energy resulting from
current along the seven separators shown in the topology map in
Figure~\ref{fig:top}. Separators $\Psi_1$ and $\Psi_2$ show a steady
increase in the free energy after $\sim$15:00~UT.  Separators $\Psi_3$
and $\Psi_7$ show no significant changes in energy from the initial
potential state; thus those curves stay near the x-axis in
Figure~\ref{fig:energy}. As seen in Figure~\ref{fig:top}, $\Psi_3$ and
$\Psi_7$ are nearly co-located and enclose many of the same domains,
so their free energies behave similarly. Separator $\Psi_4$ shows an
increase in energy until $\sim$13:00~UT, and then a decrease until the
last time. Separator $\Psi_5$ shows an increase in energy from
$\sim$18:00 UT and then a decrease. Separator $\Psi_6$ shows a larger
increase from $\sim$18:00 UT.

The mutual inductance term used in this analysis does not account for
the internal distribution of current along each separator (see
\S3.3). Separators $\Psi_3$ and $\Psi_7$ are located close together as
seen in Figure~\ref{fig:top}, but the free energies are very
small. Therefore, we assume that the mutual inductance is not
important for the overall energetics of the system.

\section{Magnetic energy buildup and coronal loop brightness
evolution}
\label{sec:results}

We have evaluated the free energy build-up along separators in AR\,10963 over a 24 hour
time period using the MCC model derived from a series of 15 times of
MDI observations. This free energy can be released through localized
heating and rapid magnetic reconnection \citep{longcope01}. We have
also introduced the brightness evolution of coronal loops observed by
XRT and {\it TRACE}. The free energy of AR\,10963 is now compared to
the observations in the Soft X-rays and EUV to determine if the free
energy is related to the coronal loop brightening.

In Figure~\ref{fig:lightcurves}, we show the light curves for four
locations selected from the coronal loop observations of {\it
Hinode}/XRT and {\it TRACE}.  Box~1 shows a brightening in XRT shortly
followed by a brightening in {\it TRACE}. Boxes~2,~3 show many
transient brightenings in both telescopes. Box~4 shows steady emission
in both XRT and {\it TRACE}. We compare these light curves to the free
energy buildup for $\sim 24$ hours as obtained by the MCC model.

We compare the
locations of separators with the coronal loops of AR\,10963 in
Figure~\ref{fig:imageNtop}, where the magnetic topology of AR\,10963
is superimposed on the XRT and {\it TRACE} images at 23:54~UT on July
14 viewed on the image plane, and the separators and light-curve box
locations are also shown.  In this section, we first investigate which
domains lie along the line of sight at various coronal heights for
each of the four light curve boxes.  Second, we discuss the coronal
loop brightness evolution in comparison with the domains and
separators.

\subsection{Domains along the line of sight}

The optically thin X-ray and EUV emission seen by XRT and {\it TRACE}
is due to contributions from plasma along the line of sight of the
observations. The MCT model is useful for determining which domains
lie along the line of sight through the corona, at various heights
above the photosphere. We determine which domains intersect the line
of sight at the locations of the four light curve boxes in order to
map which domains contribute to the observed emission.

The left panel of Figure~\ref{fig:footprints} shows the footprint of domains on the
photosphere, along with the four light curve boxes
(10$''\times$10$''$). Each footprint is bounded by spines and fan
traces, shown as black thin and dashed lines, respectively.  Each
color represents a different domain which is shown on the right side
of the Figure with the same color. The right panel of
Figure~\ref{fig:footprints} shows potential field lines with the
same colors on the left panel. The potential field lines are shown on
the XRT image as in Figure~\ref{fig:imageNtop}.

Figure~\ref{fig:los} shows the domains as a function of height along
the line of sight from each of the four boxes where the light curves
were calculated.  Each box has four points on the horizontal axis that
represent the four corners (SE, NE, SW, NW in that order) of the
$10^\arcsec\times10^\arcsec$ boxes used to calculate the light
curves. The flux domains are evaluated every 0.1\,Mm from the
photosphere.

Some domains in the left panel of Figure~\ref{fig:los} are not readily
evident in the left panel of Figure~\ref{fig:footprints}. The P2--N2 domain is not seen in
the left panel of Figure~\ref{fig:footprints} because the domain does not have a
photospheric footprint, i.e., it is a coronal domain
\citep{BeveridgeLongcope2005}. The P5--N2 domain is located between
domains P5--N3 and P5--N1. It is very thin and can be seen only near
($280^\arcsec$,~$-150^\arcsec$) in Figure~\ref{fig:footprints}, and
near 50\,Mm of Box~1 in the bottom panel in Figure~\ref{fig:los}.

We show the domains for each light curve box in the second column in
Table~\ref{tb:domains}. The domains are listed in order of coronal
height from the photosphere. For instance, the light curve in Box~1 
receives contributions from P5--N3, P5--N1 and P5--N2 below
P2--N1, although P5--N2 occupies only a very thin layer near 50\,Mm. 

To determine which domains make the main contribution to each light curve, 
we need to know the height range of the loops contributing to the light curves. 
We estimate the height of a loop by assuming that the loop is a
semicircle lying in a plane perpendicular to the solar surface, with diameter
equal to the distance between the two ends of the
loop seen by XRT and {\it TRACE} (see \S2.2 and Appendix). Using this
method, we find that the loop that clearly contributes to the light
curve in Box~1 has a height of $\sim$50\,Mm. This number takes into
account the projection due to the location of the AR\,10963 $\sim$~250$''$
solar west of central meridian.  We estimate the heights of loops that contribute
to the light curves in Box~2 -- 4 (see Figure~\ref{fig:loops} and Appendix) and find
that the estimated heights are all lower than 50\,Mm (see Table~\ref{tb:loops}). Therefore we
assume that the heights of coronal loops that produce the enhancements
in the light curves are lower than $\sim$50\,Mm. Loops in the core of
active regions tend to be relatively short, so this is a reasonable
assumption. Using this assumption, we only include domains that are lower
than $\sim$50\,Mm in the analysis.

In the high corona above $\sim$50\,Mm, the domains P4--N1, P7--N1, and
P1--N1 occupy the corona for all boxes. In addition, the domain P2--N1
extends over Boxes~1,~2,~\&3, but not Box~4. Loops contained in these
high domains would mostly be very long, as can be seen in the
potential field extrapolation in the right panel of
Figure~\ref{fig:footprints}. Assuming these loops follow RTV scaling
laws \citep{rosner78}, the density in them would be quite low because
of their long length. Thus, we assume that the domains in the corona
above the estimated height of the observed loops do not significantly
affect the enhancements in light curves. The domains below
$\sim$50\,Mm are represented with bold letters for each light curve
box in Table~\ref{tb:domains}.

\subsection{Relationships among the light curves, domains, and separators}

In this section, we investigate which separators can affect those
domains that intersect the lines of sight for the light curves of the
XRT and {\it TRACE} emission.  This result is key to investigating if
the free energy calculated by the MCC model can explain the coronal
loop brightness evolution.

Table~\ref{tb:matrix} shows the separators that can affect each domain
(see \S3.3). We show the relevant separators next to each domain in
Figure~\ref{fig:los}.  Table~\ref{tb:domains} summarizes the
relationships between the light curve boxes, domains, and
separators. The separators that can affect the domain flux are shown
in the third column in Table 3. As we described in the previous
section, we consider the domains below $\sim$50\,Mm.

\begin{figure} \epsscale{1.0}\plotone{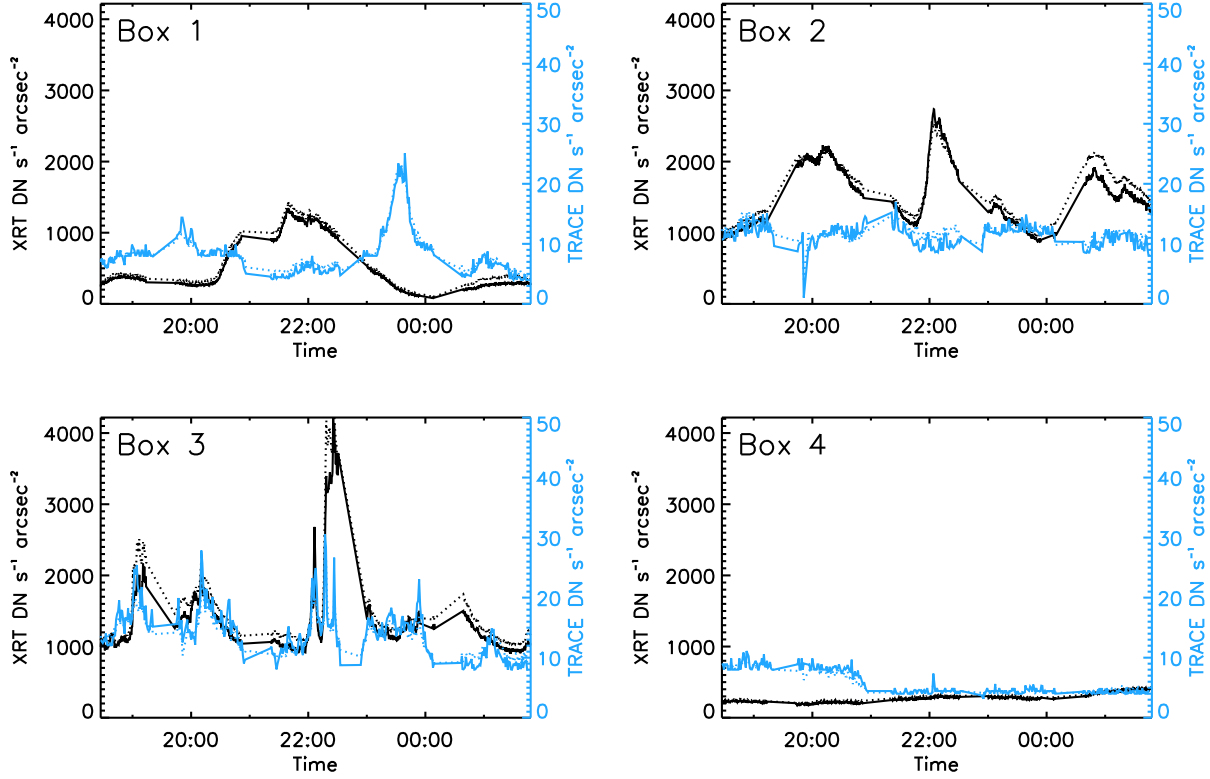}
\caption{Light curves for the four boxes in Figure
\ref{fig:xrttrace}. Black and blue lines (solid:$\sim 5'' \times 5''$,
dotted: $10'' \times 10''$) represent XRT and {\it TRACE},
respectively. Left and right axes represent the DN s$^{-1}$ arcsec$^{-2}$ for XRT and
{\it TRACE}, respectively. Left top~(Box~1):~A brightening in {\it TRACE} following
a brightening in XRT, indicating a single cooling event. Right
top~(Box~2):~Transient brightenings in XRT and only very small changes in
{\it TRACE}. Left bottom~(Box~3):~Transient brightenings in both XRT and
{\it TRACE}. Right bottom~(Box~4):~Steady emission in both XRT and {\it TRACE}.}
\label{fig:lightcurves} 
\end{figure}

\begin{figure} \epsscale{0.7}\plotone{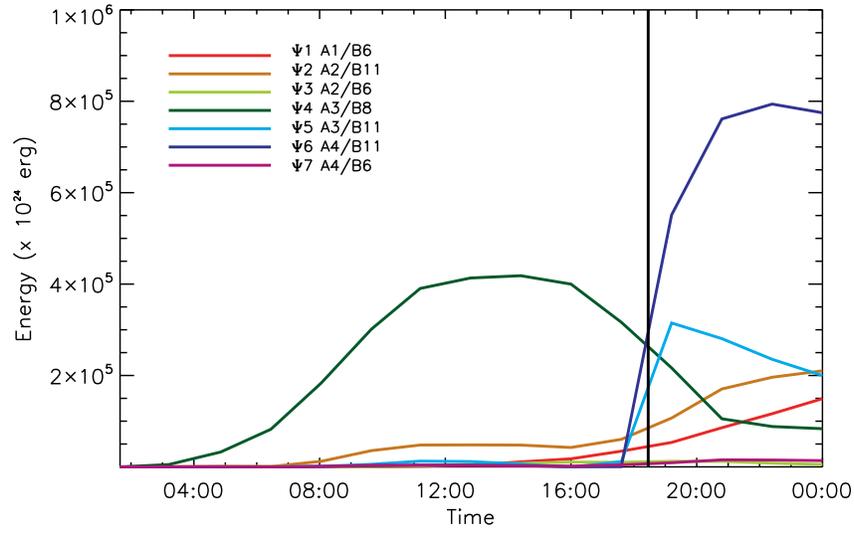}
\caption{Energy buildup from 01:39\,UT on July 14 (current=0 at this
  time) for each separator. Colors are the same as in Figure~\ref{fig:top}. Black
thick line represents the start time of the light curves in
Figure~\ref{fig:lightcurves}. $\Psi_1$ and $\Psi_2$:~A steady increase
after $\sim$15:00\,UT. $\Psi_3$ and $\Psi_7$:~no significant changes
in energy from the initial potential state. $\Psi_4$:~an increase
until $\sim$13:00\,UT and then a decrease. $\Psi_5$:~an increase from
$\sim$18:00\,UT and then a decrease. $\Psi_6$:~an increase from
$\sim$18:00\,UT.}
\label{fig:energy}
\end{figure}

\begin{figure} 
\epsscale{1.2}\plottwo{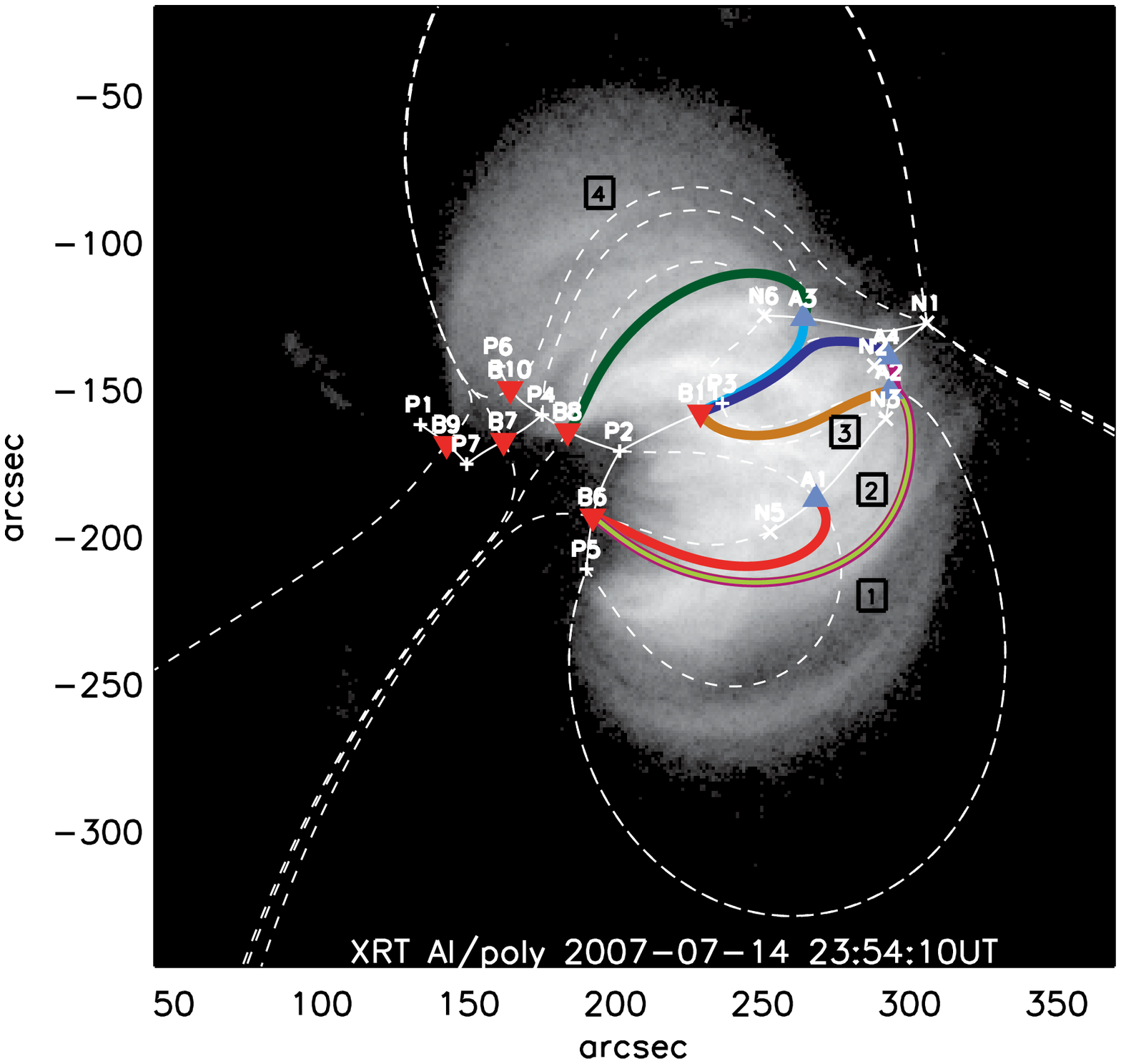}{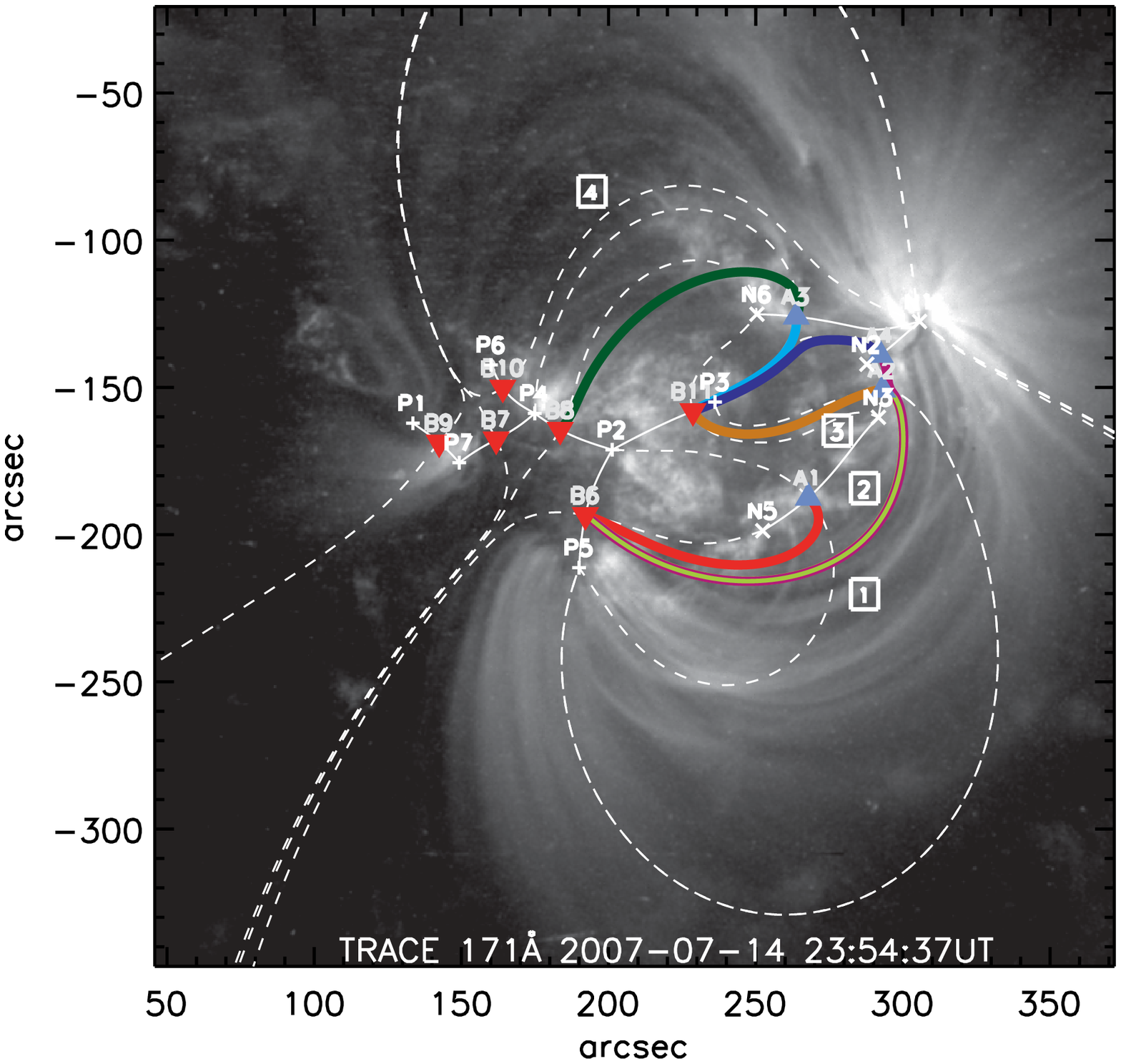}
\caption{Locations of separators on XRT and {\it TRACE} observations. The
symbols are the same as in Figure~\ref{fig:top}. Fan traces (white
dashed lines), spines (white thin solid lines), and separators
(colored lines) are placed on the view from the Earth (image
plane). The separator colors are the same as in
Figures~\ref{fig:top},~\ref{fig:energy}. The locations of separators
are slightly shifted comparing with the locations in
Figure~\ref{fig:top} due to the difference between the image plane and the
tangent plane. Light curve boxes are also represented as four boxes.}
\label{fig:imageNtop}
\end{figure}

\begin{figure}
\epsscale{1.2}\plottwo{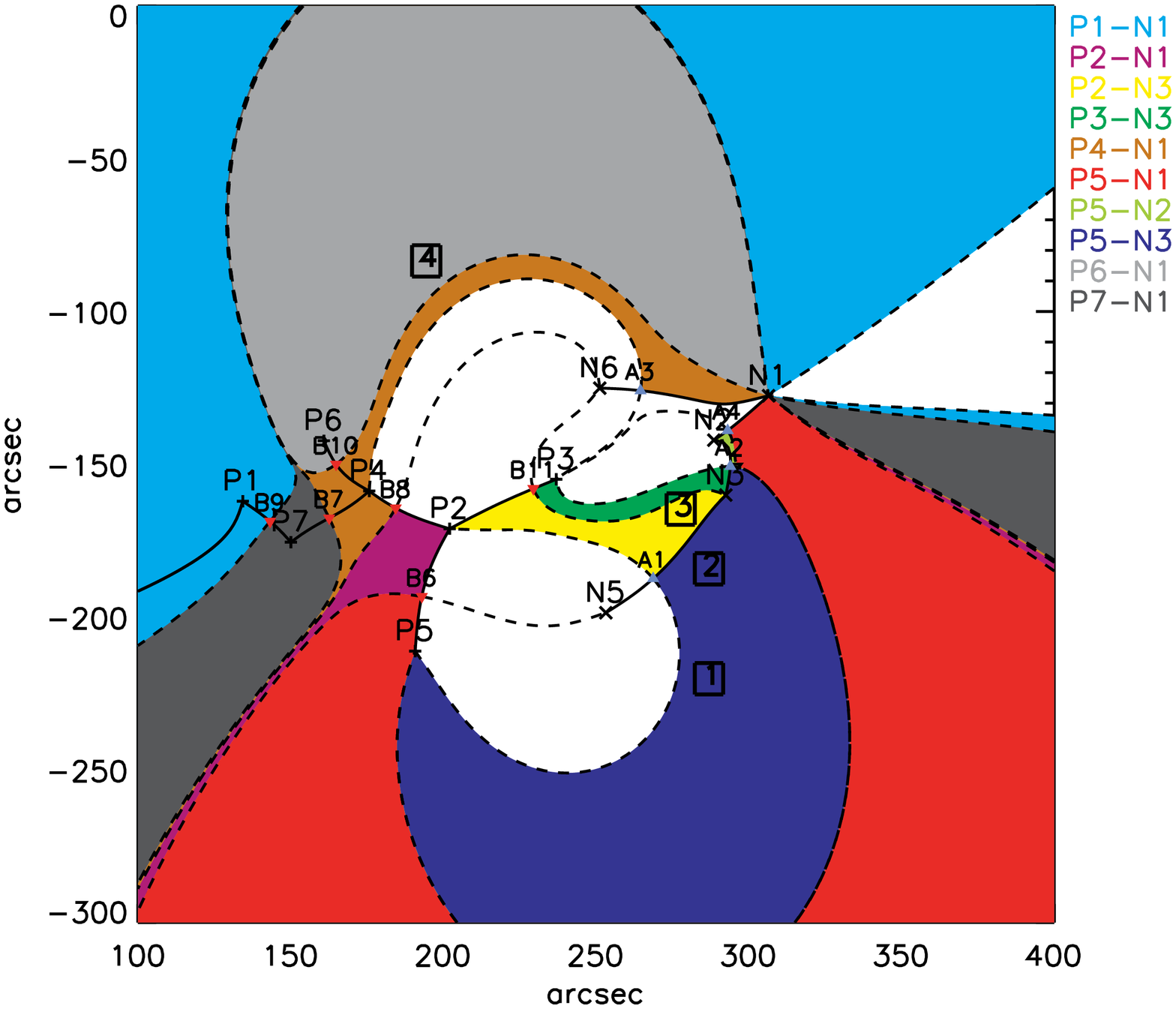}{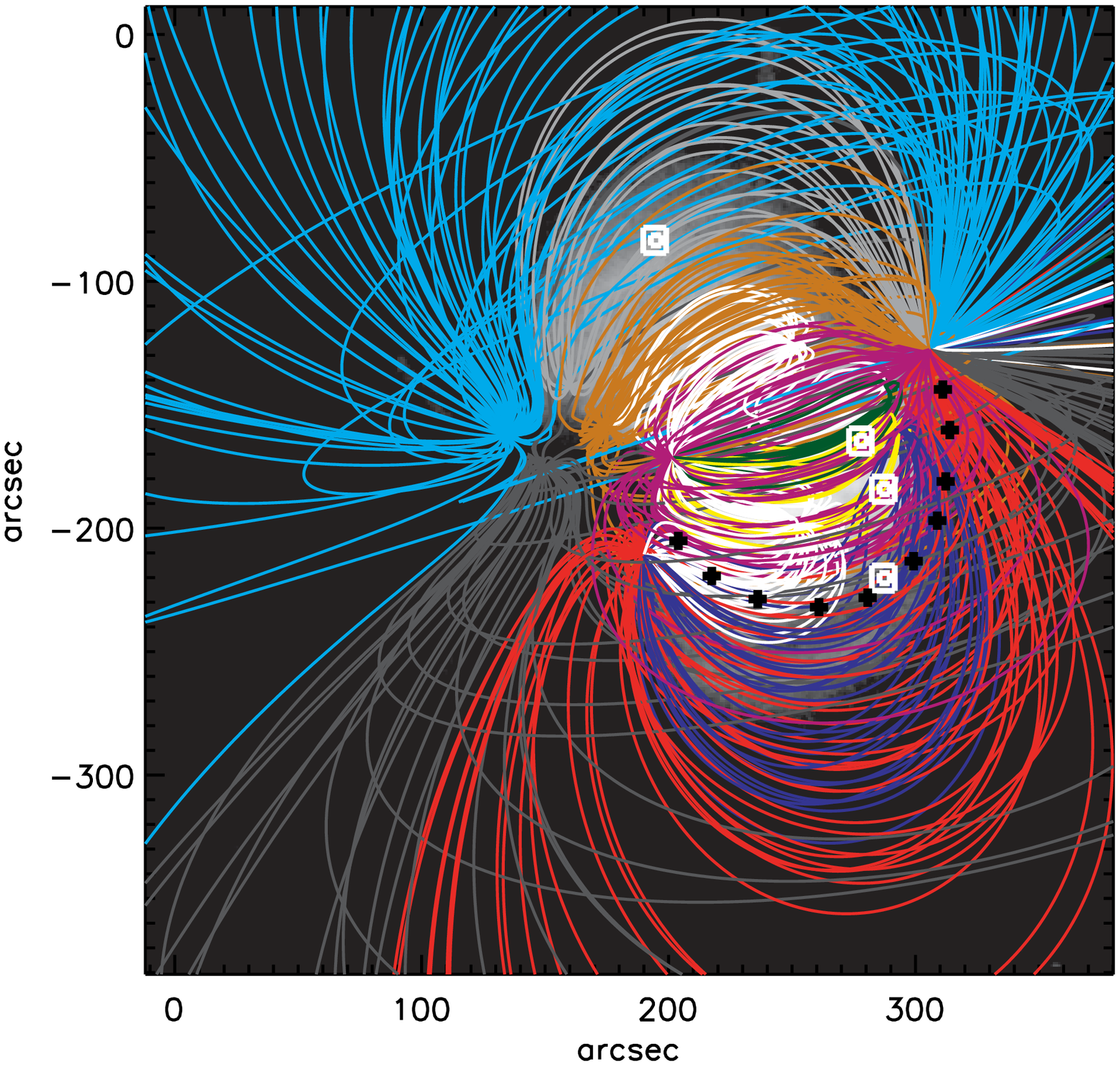}
\caption{Left: Domain footprints on the image plane. Symbols are the same as in
Figures~\ref{fig:top},~\ref{fig:imageNtop}. The domain footprints are divided
by the spines (black solid lines) and fan traces (black dashed lines) on the
photosphere. Domains present along the lines of sight for the light curve boxes
are shown in color. The positive and negative sources of the domains are
labeled on the right side of the figure. Right: Potential field
lines of each domain with the same colors as in the left panel. White boxes are the light
curve boxes. The loops in Figure~\ref{fig:xrttrace} are also
represented with $''+''$.} 
\label{fig:footprints}
\end{figure}

\begin{figure}
\epsscale{0.8}\plotone{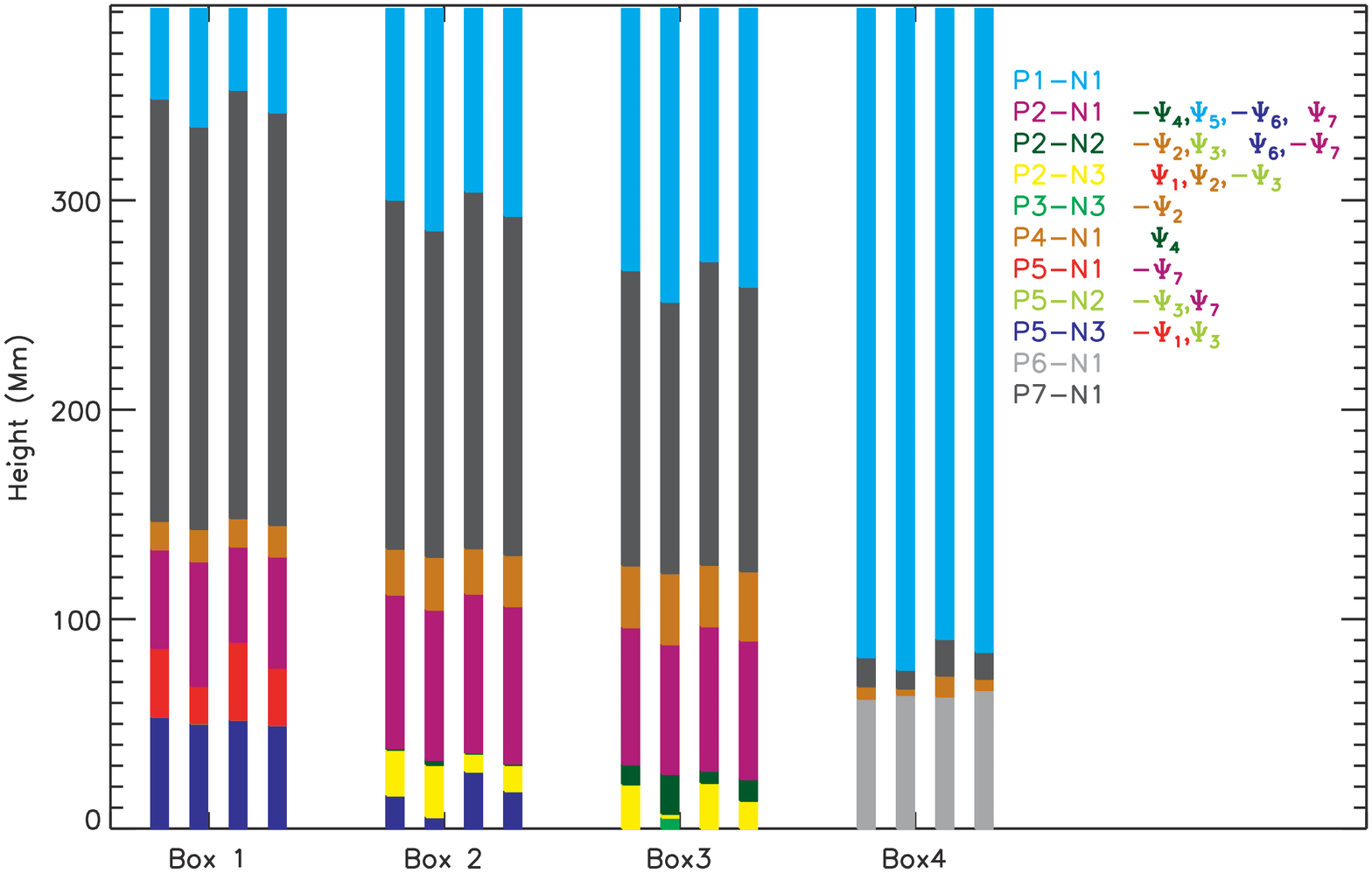}
\epsscale{0.8}\plotone{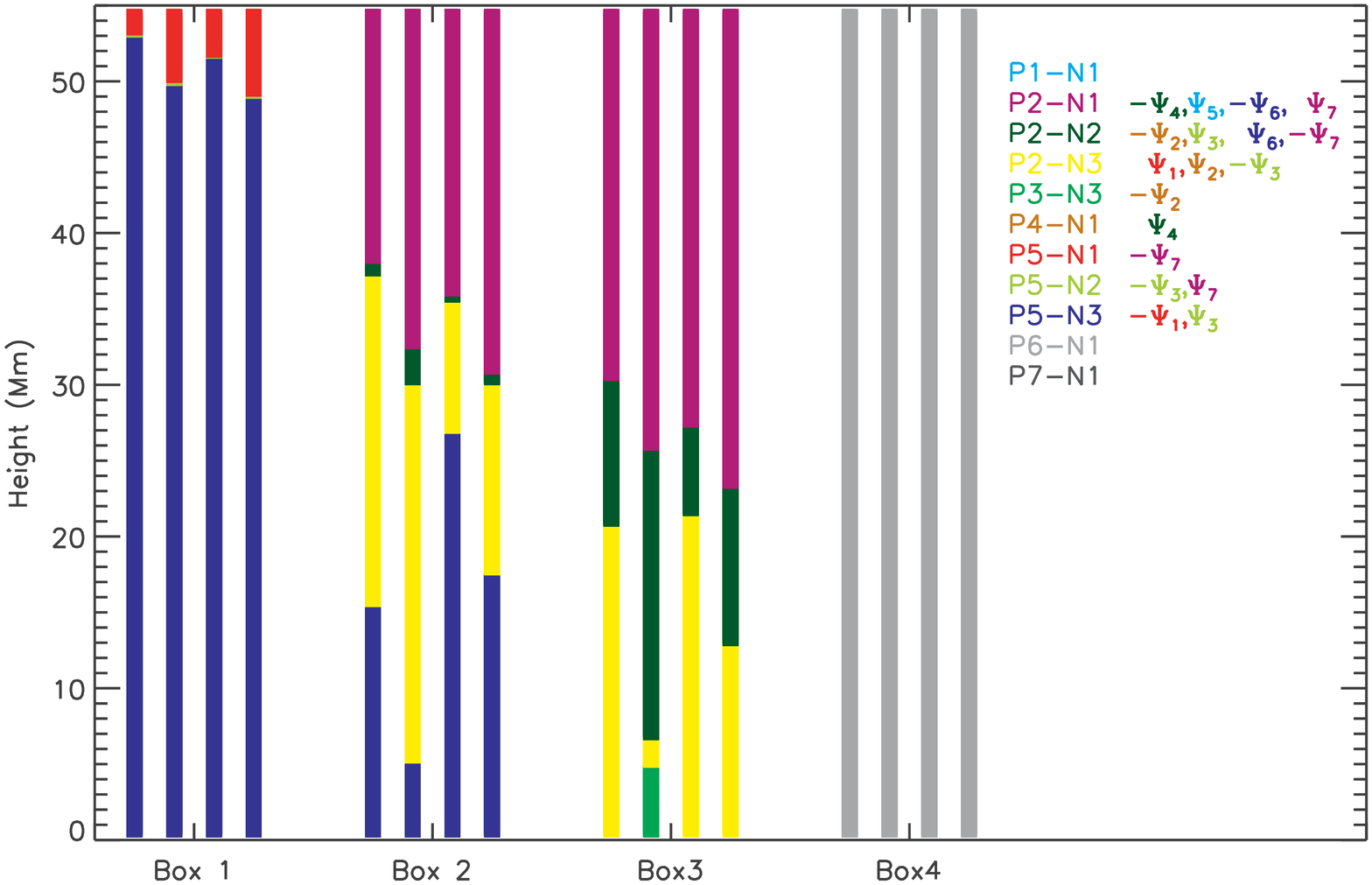}
\caption{Domains along the line of sight through the corona at the locations of
the light curve boxes.  The top and bottom panels show the domains over a large
height range and a small height range, respectively. The domains are shown with
the same colors as in Figure~\ref{fig:footprints}. Each box has four points in
the horizontal direction that represent the four corners (SE, NE, SW, NW in
order) of $\sim 10'' \times 10''$ square of the boxes. The separators that
could affect the domain fluxes are represented on the right of the domains with
the same colors as in Figure~\ref{fig:imageNtop}.} 
\label{fig:los}
\end{figure}

The light curves of Box~1 show a transient brightening in the {\it
TRACE} observation after a decrease of intensity in the XRT
observation, implying that there is a cooling of the coronal loop from
X-ray temperatures ( $\ga$ 2\,MK) through EUV temperatures ($\sim$
1\,MK, see Figure~\ref{fig:lightcurves}).  Observations of a coronal
loop brightening in EUV following a loop brightening in X-rays have
been examined previously. Many authors have studied the correlation
between X-ray and EUV brightenings and the implications of this
correlation for the loop heating mechanism \citep{wineb05, urra09,
ugarte06, warren09a}. \citet{wineb05} study several loops that are
relatively seclude and uniquely shaped so that they are able to follow
the same loop as it cools through the different filters. They conclude
that the EUV emission can be produced by a loop that emits in the
X-rays several hours before, although the delay time is difficult to
reproduce through hydrodynamic modeling.

We have evaluated an approximate cooling time of the loop including
the light curve Box~1 in Figure~\ref{fig:xrttrace}. Assuming that
plasma is at a temperature of 10 MK when it peaks in XRT, and is at 1
MK when it peaks in TRACE, and the loop is semicircular with a radius
of 50$''$, width of 10$''$ and has a uniform cross section.  With
these assumptions, the density at the time of peak intensity in TRACE
is about $2 \times 10^9 {\rm cm}^{-3}$, and the loop has a cooling time on
the order of about an hour to go from 10 MK to 1 MK. The temperature
and density evolution are calculated using an analytic solution to the
1D hydrodynamic energy equation \citep{cargill95}, and the assumptions
are not to be believed in detail. However this calculation is a
reasonable zeroth order estimate for the cooling time. Comparing the
estimated cooling time with the light curve of Box~1
(Figure~\ref{fig:lightcurves}) that shows about 2 hours between peak
in XRT and TRACE, it is likely that the loop is cooling from XRT to
TRACE temperatures. Thus we conclude that the emission in Box~1 is
probably due to a cooling loop.

The light curves in Box~1 are integrated over domains P5--N3, P5--N2,
and P5--N1 below $\sim$50\,Mm. These domains are related to the
separators $\Psi_1, \Psi_3$, and $\Psi_7$. The latter two separators,
$\Psi_3$ and $\Psi_7$, have relatively little free energy with very
little change as shown in Figure~\ref{fig:energy}.  Although the
domain P5--N3 associated with $\Psi_1$ is along the line of sight of
Box~1, the geometry of this separator indicates that it is probably
not a key component of the emission in this area. In the right panel
of Figure~\ref{fig:footprints}, the loop that is included in Box 1 is
located in a region dominated by the magnetic field lines of domains
P5–-N3 and P5–-N1. In addition, the height of $\Psi_3$ and $\Psi_7$
are both 50\,Mm, and their lengths are both about 160\,Mm. In
contrast, $\Psi_1$ is much lower and shorter, with a height of 27\,Mm
and a length of 89\,Mm.  A few studies have found that longer coronal
loops lead to longer delay times between the appearance of the X-ray
loop and the appearance of the EUV loop \citep{ugarte06, asch03}. The
delay time between the XRT and {\it TRACE} emission peaks in Box~1 is
relatively long, thus the emission is likely to have come from longer
loops that are located higher in the corona. The separators $\Psi_3$
and $\Psi_7$ are more likely associated with long loops than $\Psi_1$,
so the emission in Box~1 is probably related to separators $\Psi_3$
and $\Psi_7$.

Therefore, our analysis shows that the emission variation in the area
of Box~1 as summarized by its light curves most likely comes from the
domains associated with the separators $\Psi_3$ and $\Psi_7$, which
have a small free energy. Thus it is possible that the loop is in a
nearly potential environment, so that not much energy release can
occur. This small free energy may be the reason why there are not many
repeated transient events in Box~1, as there are in Boxes~2 and 3.

Boxes~2 and 3 are related to domains P2--N3, P2--N2, P3--N3, and
P2--N1. These domains are related to separators $\Psi_1, \Psi_2,
\Psi_3, \Psi_4, \Psi_6$, and $\Psi_7$. Boxes~2 and 3 display several
transient brightenings in Figure~\ref{fig:lightcurves}. Thus it is
likely that the variable emission in these areas comes from the
domains associated with a high number of separators, including, but
not limited to, the separators for Box~1.  Additionally the free
energy associated with $\Psi_1$, $\Psi_2$, and $\Psi_6$ increases over
the time period of the observations, indicating that the fields
associated with these separators are stressed and non-potential. This
configuration could provide an environment for continuous impulsive
events in Boxes~2 and 3 through multiple reconnection processes
occurring at the separators.

Recent studies have shown that continuous impulsive heating events are
common in loops located in the core of an active region \citep{urra09,
warren07}.  \citet{warren07} presented bi-directional flows as
evidence of magnetic reconnection during the evolution of coronal
loops in the XRT and the EIS observations. The loops observed by
\citet{wineb05} shift positions as they cool, which may be evidence
for a previous reconnection event. Therefore, our conclusion that
areas of active regions that show impulsive emission are probably
related to reconnection events along separators is broadly consistent
with previous observations.

Transient brightenings in this active region during the observation
period have different characteristics.  The light curves of Box~3 show
transient brightening in both XRT and {\it TRACE}.  However, the light
curves for Box~2 show transient brightenings in the XRT, but the {\it
TRACE} light curve shows only very small changes. This situation may
indicate that loops in Box~2 are being heated more frequently and have
insufficient time to cool into the {\it TRACE} passband before being
reheated. This kind of difference can not be addressed by our analysis
since we only locate the possible sites for reconnection, and do not
determine the frequency with which reconnection takes place.

The light curves of Box~4, which shows steady emission in both XRT and
{\it TRACE}, intersect domain P6--N1 below $\sim$50\,Mm. This domain
does not have any associated separators. ``Steady heating'' has been
suggested as the heating mechanism for loops that have {\it TRACE}
moss at their footpoints, as the loop in Box~4 does
\citep{antiochos03,warren08}. In this analysis, the emission of Box~4
comes from a domain that is not associated with any separators,
indicating that larger-scale reconnection events are unlikely to be
contributing to the heating of this plasma. Thus the steady heating
proposed by \citet{antiochos03} is not likely to be due to
reconnection events at separators shown in the larger-scale magnetic
field configuration. Rather, the source of the heating that results in
the emission observed in Box~4 may be due to many small reconnection events
or nanoflares from local field tangling \citep[e.g.][]{antiochos03,
parker83a, parker83b, parker88, klimchuk06}, or alternatively, MHD waves
\citep[e.g.][]{kumar2006,offman2008,antolin2010}.

Several studies for the heating mechanism of coronal loops have
difficulty explaining the brightness of observations in both EUV and
X-ray emission and the delay between the appearance of X-ray and EUV
loops \citep{warren06, ugarte06}. However, recent studies have pointed
out the importance of magnetic properties to explain the coronal loop
evolutions \citep{urra09, warren09a}. Our result also concludes that
the magnetic topology plays a significant role for the study of the
heating mechanism of non-flaring coronal loops.

\section{Conclusion}
\label{sec:conclusion}

In this analysis, we examined {\it Hinode}/XRT and {\it TRACE}
observations of particularly well-resolved coronal loops in an active
region that show different characteristics in their emission
properties. In order to understand these differences, we compare the
coronal loop evolution represented by four light curves from XRT and
{\it TRACE} observations to the free energy calculated by the Minimum
Current Corona (MCC) model. We compare the free energy gained from the
motions of the magnetic footpoints with emission patterns that include
steady emission and transient emission. This analysis marks the first
time that the MCC model has been applied and tested as a full time
series analysis for a quiet active region to investigate coronal loop
evolution in plasma of different temperatures.

This work shows that the magnetic topology of an active region can
provide crucial information for understanding the evolution of coronal
loop brightness. One of the major questions in coronal loop evolution
studies is the difference in the heating mechanism for steady emission
versus transient emission. We find that regions which have very
dynamic emission are related to a high number of separators that have
enhanced free energy, indicating that reconnections along these
separators may play a role in repeated transient heating events. In
contrast, the light curve in Box~1 that shows a single transient
brightening in the XRT observations followed by a brightening in {\it
TRACE} is related to separators with little free energy. This
relationship indicates that the fields in the vicinity of Box~1 are
nearly potential to begin with. Thus there is not a lot of excess
energy available for release through repeated reconnection events in
this region, and repeated transient events are not observed in the
light curve.

Finally, the steady emission in both XRT and {\it TRACE} comes from
the domains that are not associated with any separator, indicating
that reconnections involving the larger-scale field configuration are
probably not associated with steady heating, and implicating either
nanoflares or MHD waves as a possible heating source for these
loops. We have found a topological explanation for the two
classes of coronal loops, those that emit steadily and those that show
transient emission properties.

The {\it Solar Dynamics Observatory} ({\it SDO}) launched on 2010
Feb.~11, will undoubtedly shed new light on the problem of active
region loop heating. The Helioseismic and Magnetic Imager (HMI)
onboard {\it SDO} will provide a high cadence of vector magnetogram
data. These observations will provide detailed temporal information of
the photospheric boundary evolution with which to place in context the
observations by the Atmospheric Imaging Assembly (AIA) and its EUV
observations in various coronal temperatures. The data from this
instrument suite will complement the analysis available, as
demonstrated here, from the Magnetic Charge Topology and Minimum
Coronal Current models for further insights into the mechanics of
active region heating.

\acknowledgments

We thank Bish Ishibashi for the helpful comments on this paper. We
thank the anonymous referee for his/her comments that improved this
paper. This work was supported by NASA grant NNM07AA02C to the Smithsonian
Astrophysical Observatory and by the Air Force Office of Scientific
Research contract FA9550-06-C-0019 to NorthWest Research
Associates. {\it Hinode} is a Japanese mission developed and launched
by ISAS/JAXA, collaborating with NAOJ as a domestic partner, NASA and
STFC (UK) as international partners. Scientific operation of the {\it
Hinode} mission is conducted by the {\it Hinode} science team
organized at ISAS/JAXA. This team mainly consists of scientists from
institutes in the partner countries. Support for the post-launch
operation is provided by JAXA and NAOJ (Japan), STFC (U.K.), NASA,
ESA, and NSC (Norway).

\appendix

\section{Appendix}
\label{sec:appendix}

\begin{figure} 
 \begin{tabular}{cccc}
    {\epsscale{0.30}\plotone{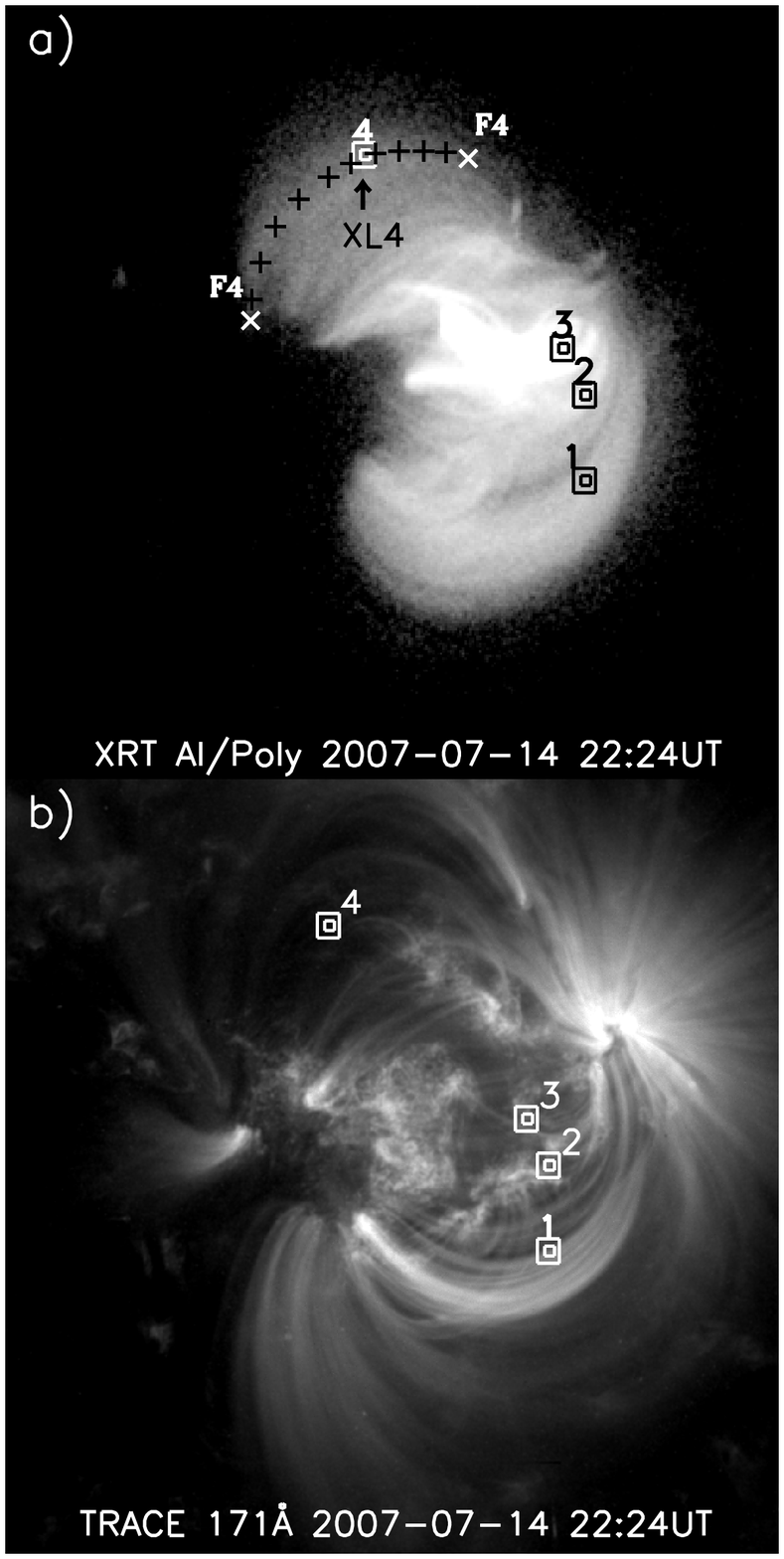}} \hspace{-1.4cm}
    {\epsscale{0.30}\plotone{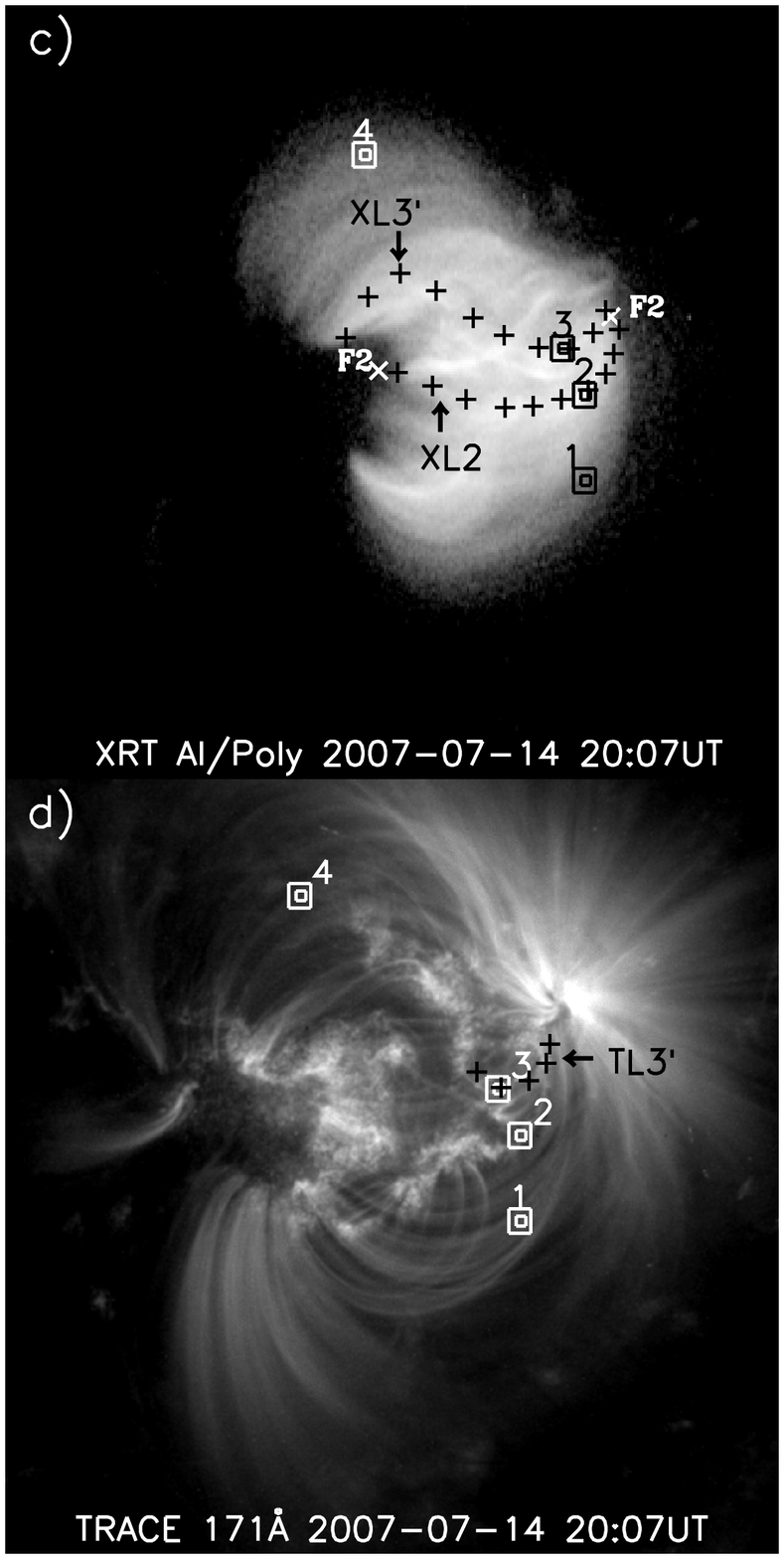}} \hspace{-1.4cm}
    {\epsscale{0.30}\plotone{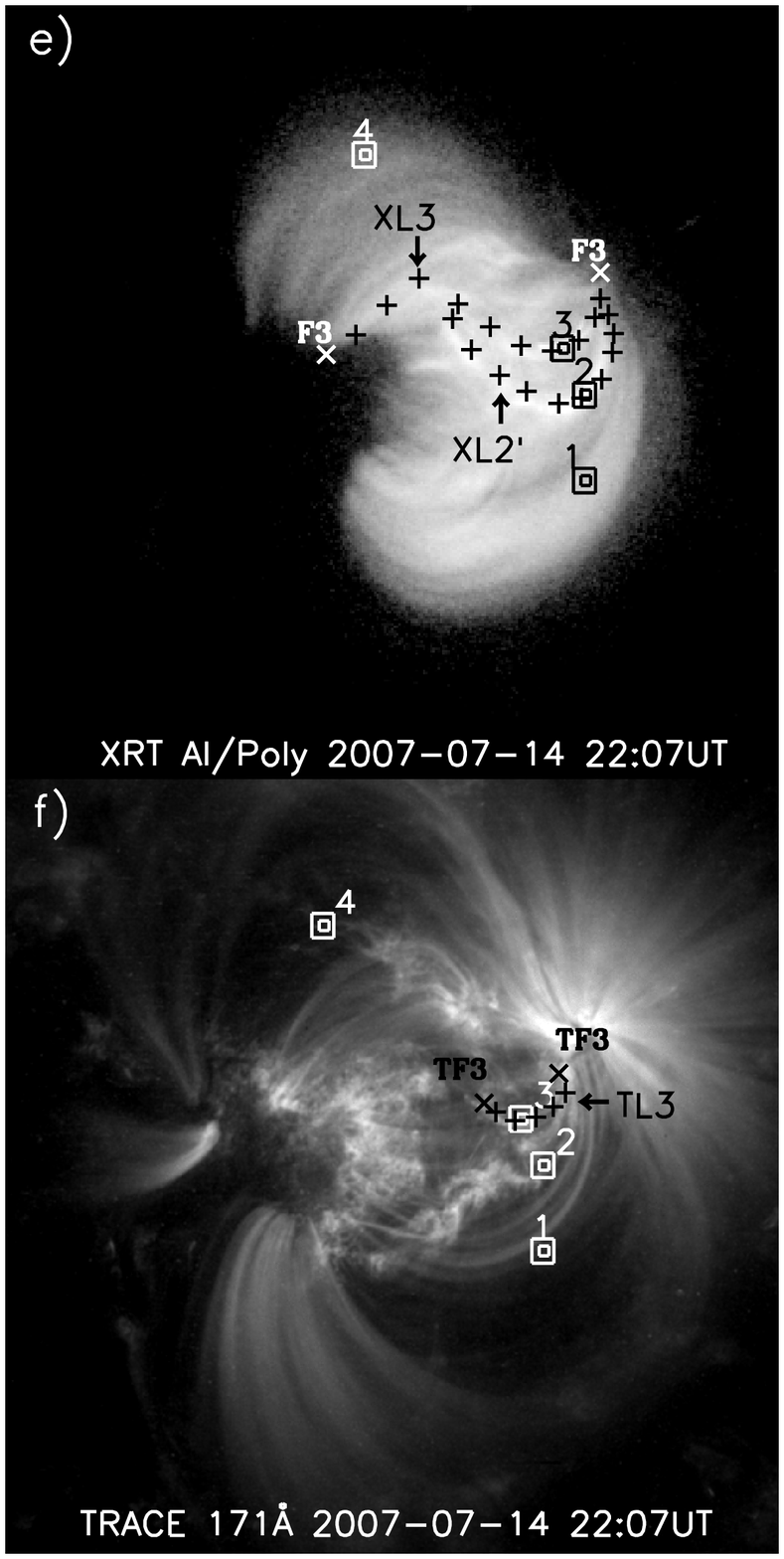}} \hspace{-1.4cm}
    {\epsscale{0.30}\plotone{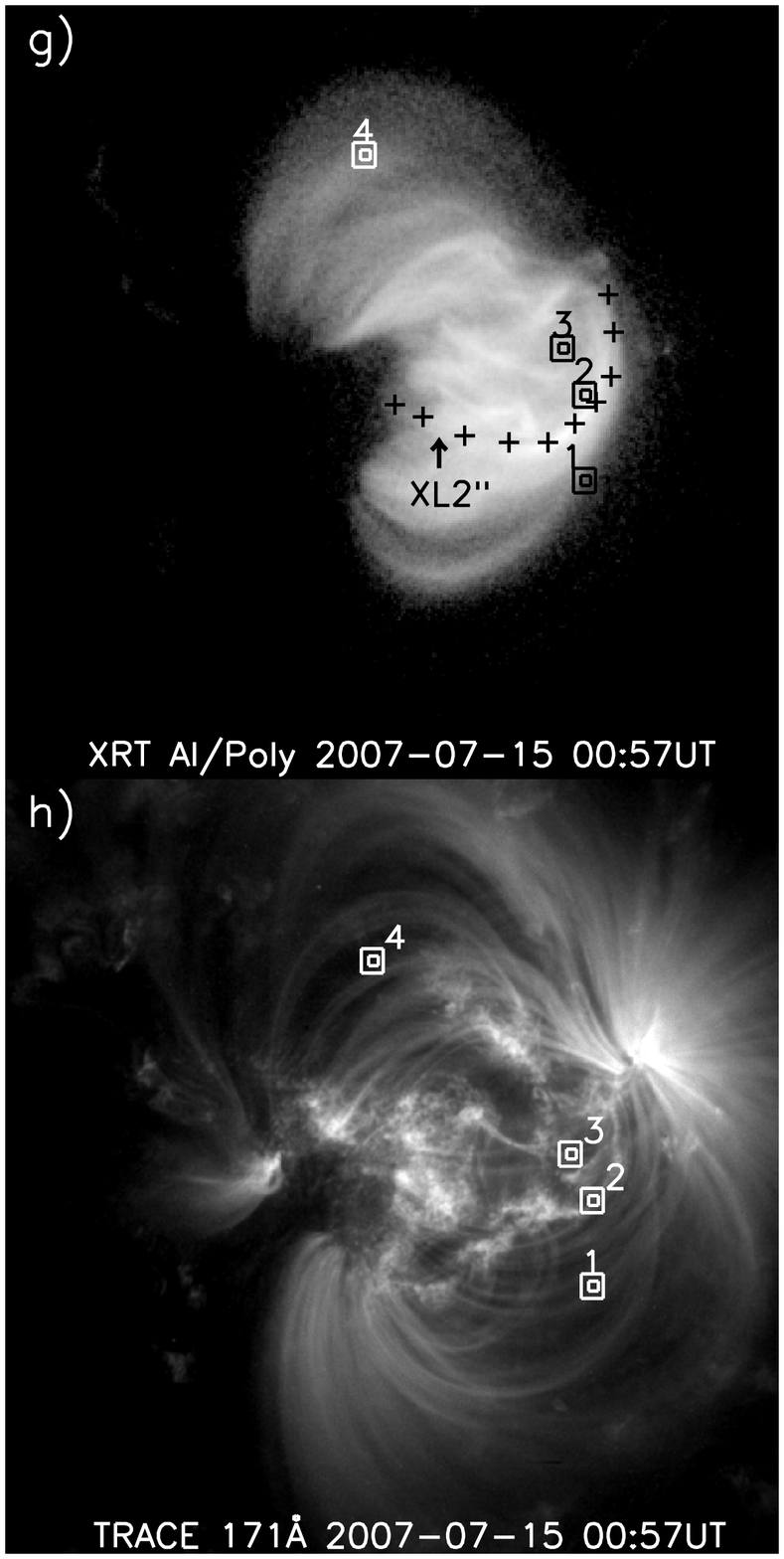}} \hspace{-1.4cm}
 \end{tabular}
\caption{Loops that produce the enhancements in light curves are
represented by $''+''$ symbols along the loop. a) XRT image for loops
that contribute to the light curve in Box~4 b) Corresponding TRACE
image. c)--h) Loops that produce the enhancements of the light curves
Box~2 and Box~3 at three different times in
Figure~\ref{fig:lightcurves}. The ends of loops are shown as a symbol
$''\times''$ with letters, F2, F3, F4, and TF3 for several loops,
which are applied to estimate the heights of the loops (see
Table~\ref{tb:loops}). Light curve boxes are also shown as four
boxes.}
\label{fig:loops}
\end{figure}

Figure~\ref{fig:loops} shows several loops and their ends that produce
enhancements in light curves. Using these loops, we estimate the
height of the loops that contribute to the light curves (see \S4.1).

Panel a) shows structures (XL4) that contribute to the light curves in
Box~4. The region that contributes to the light curves in Box~4 is a
diffuse area and does not show as a single loop, and there are no
changes in X-ray and EUV emission. The location of the structure for
the Box~4 is not represented in panel b) because the loop is not seen
on the {\it TRACE} observations. An estimated height of loops using
the ends of loops (F4) is represented in Table~\ref{tb:loops} (see
also \S4.1).

The transient brightenings in the light curve Box~2 and Box~3 are
produced by several different loops at different times. Therefore, in
panels c)-h), we show the loops that produce enhancements in the light
curves of Box~2 and Box~3 at three different times when the transient
brightenings are observed by XRT and/or {\it TRACE}.  In panel c),
each loop appearing in Box~2~(XL2) and Box~3~(XL3$'$) that shows
enhancements in the light curve in the XRT observation at around
20:00\,UT is represented. The ends of loop seen by XRT for Box~2 is
marked by F2 in panel c). The loop (TL3$'$) that shows enhancements in
the {\it TRACE} light curve in Box~3 is a small loop shown in panel
d). In panels e) and f), we show the loops that correspond to
enhancements at around 22:00\,UT. The ends of loop seen by XRT and
{\it TRACE} for Box~3 are marked with F3 and TF3 in panels e) and f),
respectively. In panel e), the loop (XL2$'$) that contributes to the
light curve Box~2 at that time is also shown. In panel g), the loop
(XL2$''$) that shows the enhancement in the light curve for Box~2 at
around 01:00\,UT is shown. At this time, there is no distinct loop in
Box~3 in the XRT observation, although there is a small enhancement in
the light curve. In panel h), no loops are represented because there
is no enhancement in the light curves and brightening in the TRACE
observations at this time. The loops appearing in Box~2 and Box~3 at
the last three times in Figure~\ref{fig:loops} look like different
loops (see animation movie), though their locations are similar.

We represent the height of loops, L1, XL2, XL3, TL3, and XL4, in
Table~\ref{tb:loops}. The distance between the ends of loops, XL3$'$,
TL3$'$, XL2$'$, and XL2$''$, are similar or smaller than the loops in
Table~\ref{tb:loops}. Therefore, the heights of these loops are lower
than the heights of the loops in Table~\ref{tb:loops}.

\bibliographystyle{apj} \bibliography{ms}

\clearpage

\begin{deluxetable}{cllll} \tabletypesize{\scriptsize}
\tablecaption{Separator properties} \tablewidth{0pt}
\tablehead{ \colhead{Separators} & \colhead{Nulls} & \colhead{Length
(Mm)} & \colhead{Height (Mm)} & \colhead{Domain} } 

\startdata $\Psi_{1}$ & A1/B6 & 89 & 27 & (P5--N5) \\ $\Psi_{2}$
& A2/B11 & 55 & 8 & -(P3--N3)\tablenotemark{*} \\ $\Psi_{3}$ & A2/B6 &
153 & 50 & (P5--N5)+(P5--N3)  \\ $\Psi_{4}$ & A3/B8 & 104 &
22 & -(P4--N6)\tablenotemark{*} \\ $\Psi_{5}$ & A3/B11 & 44 & 11 & (P3--N6) 
\\ $\Psi_{6}$ & A4/B11 & 61 & 11 & (P3--N6)+(P3--N1) \\
$\Psi_{7}$ & A4/B6 & 160 & 50 & (P5--N5)+(P5--N3)+(P5--N2) \\
\enddata \tablenotetext{*}{`$-$' represents an opposite direction in
flux flows between flux domain and separator}
\label{tb:separators}
\end{deluxetable}

\begin{deluxetable}{ccccccccc} \tabletypesize{\scriptsize}
\tablecaption{Connectivity matrix in terms of source and separator fluxes} 
\tablewidth{0pt} \tablehead{
\colhead{ } & \colhead{P$_\infty$} & \colhead{P1} & \colhead{P2} &
\colhead{P3} & \colhead{P4} & \colhead{P5} & \colhead{P6} &
\colhead{P7} }

\startdata N1 & $\Phi_{P_{\infty}}$ & $\Phi_{P1}$ & \tablenotemark{*} &
$\Psi_6-\Psi_5$ &$\Phi_{P4}+\Psi_4$ & $\Phi_{P5}-\Psi_7$ & $\Phi_{P6}$
& $\Phi_{P7}$ \\ N2 & 0 & 0 &
\tiny$\Phi_{N2}-\Phi_{P3}+\Psi_6-\Psi_2-\Psi_7+\Psi_3$ &
$\Phi_{P3}-\Psi_6+\Psi_2$& 0 &$\Psi_7-\Psi_3$ & 0 & 0 \\ N3 & 0 & 0 &
$\Phi_{N3}+\Psi_2-\Psi_3+\Psi_1$ & $-\Psi_2$ & 0 & $\Psi_3-\Psi_1$ & 0
& 0 \\ 
N5 & 0 & 0 &
$\Phi_{N5}-\Psi_1$ & 0 & 0 &$\Psi_1$ & 0 & 0 \\ N6 & 0 & 0 &
$\Phi_{N6}-\Psi_5+\Psi_4 $ & $\Psi_5$ & $-\Psi_4$ & 0 & 0 & 0 \\
\enddata \tablecomments{ $P_{\infty}$ is a positive source ``at infinity'' 
introduced to balance the flux from active region. The flux of
$P_{\infty}$ is about 10 \% of total unsigned flux.}
\tablenotetext{*}{$\Phi_{N1}-\Phi_{P_{\infty}}-\Phi_{P1}-\Phi_{P6}-\Phi_{P7}-\Psi_6+\Psi_5-\Phi_{P4}-\Psi_4-\Phi_{P5}+\Psi_7$}
\label{tb:matrix}
\end{deluxetable}

\begin{deluxetable}{cll} \tabletypesize{\scriptsize}
\tablecaption{Domains and separators that influence light curve boxes}
\tablewidth{0pt} \tablehead{ \colhead{Boxes} & \colhead{Flux domains}
& Separators } \startdata Box 1 & \textbf{P5--(N3, N2, N1)}, P2--N1,
P4--N1, P7--N1, P1--N1 & $\Psi_1, \Psi_3, \Psi_7$ \\ Box 2 & \textbf{P5--N3,
P2--(N3, N2, N1)}, P4--N1, P7--N1, P1--N1 & $\Psi_1, \Psi_2, \Psi_3,
\Psi_4, \Psi_6,
\Psi_7$ \\ Box 3 & \textbf{P3--N3, P2--(N3, N2, N1)}, P4--N1, P7--N1, P1--N1 &
$\Psi_1, \Psi_2, \Psi_3, \Psi_4, \Psi_6, \Psi_7$ \\ Box 4 & \textbf{P6--N1},
P4--N1, P7--N1, P1--N1 & \\
\enddata \tablecomments{\textbf{Bold} represents the domain that
used to relate separators. }
\label{tb:domains}
\end{deluxetable}

\begin{deluxetable}{llll} \tabletypesize{\scriptsize}
\tablecaption{Loops and estimated heights}
\tablewidth{0pt} \tablehead{ \colhead{Loops} & \colhead{Time}
& Height (Mm) & Note} 
\startdata L1 & 22:24 UT$^*$& 50 & Figure~\ref{fig:xrttrace} \\
XL2 & 20:07 UT & 40 & Panel c) in Figure~\ref{fig:loops} \\
XL3 & 22:07 UT & 48 & Panel e) in Figure~\ref{fig:loops} \\
TL3 & 22:07 UT & 14 & Panels f) in Figure~\ref{fig:loops} \\
XL4 & 22:24 UT & 46 & Panels a) in Figure~\ref{fig:loops} \\
\enddata \tablecomments{`$^*$' represents a time when the XRT
  observation shows a brightening.}
\label{tb:loops}
\end{deluxetable}

\end{document}